\documentclass[aps,  preprint,  twocolumn,  10pt,  nofootinbib]{revtex4}
\usepackage{hyperref}
\usepackage{tensor}
\usepackage{amsfonts}
\usepackage{amsmath}
\usepackage{cancel}
 
\usepackage{amssymb}
\usepackage{graphicx}%
\usepackage{bigints}
\newcommand{\be}{\begin{equation}}
\newcommand{\ee}{\end{equation}}
\newcommand{\Be}{\begin{eqnarray}}
\newcommand{\Ee}{\end{eqnarray}}

\usepackage{graphicx}
\usepackage{subfigure}
\usepackage{epsf}
\usepackage{bm}
\usepackage{amsmath}
\usepackage{amsfonts}
\usepackage{amssymb}
\usepackage{graphicx}
\usepackage{tabularx}
\usepackage{color}%
\providecommand{\U}[1]{\protect\rule{.1in}{.1in}}

\newcommand{\ie}{\begin{equation}}
\newcommand{\fe}{\end{equation}}

\newcommand{\mincir}{\raise
-3.truept\hbox{\rlap{\hbox{$\sim$}}\raise4.truept\hbox{$<$}\ }}
\newcommand{\magcir}{\raise
-3.truept\hbox{\rlap{\hbox{$\sim$}}\raise4.truept\hbox{$>$}\ }}

\providecommand{\U}[1]{\protect\rule{.1in}{.1in}}

\usepackage{tikz,  xcolor,  hyperref}

\definecolor{lime}{HTML}{A6CE39}
\DeclareRobustCommand{\orcidicon}{%
	\begin{tikzpicture}
	\draw[lime, fill=lime] (0,  0) 
	circle [radius=0.16] 
	node[white] {{\fontfamily{qag}\selectfont \tiny ID}};
	\draw[white, fill=white] (-0.0625,  0.095) 
	circle [radius=0.007];
	\end{tikzpicture}
	\hspace{-2mm}
}

\foreach \x in {A, ..., Z}{%
	\expandafter\xdef\csname orcid\x\endcsname{\noexpand\href{https://orcid.org/\csname orcidauthor\x\endcsname}{\noexpand\orcidicon}}
}

\begin{document}

\title{Charged Black Holes with Yukawa Potential}

\author{A. A. Ara\'{u}jo Filho $^{1,  2}$ \orcidB{}}
\email{dilto@fisica.ufc.br}
\author{Kimet Jusufi $^{3}$\orcidA{}}
\email{kimet.jusufi@unite.edu.mk}
\author{B. Cuadros--Melgar$^{4}$\orcidC{}}
\email{bertha@usp.br}
\author{Genly Leon$^{5,  6}$\orcidD{}}
\email{genly.leon@ucn.cl}
\author{Abdul Jawad$^{7,8}$\orcidE{}}
\email{abduljawad@cuilahore.edu.pk}
\author{C.E. Pellicer$^9$\orcidF{}} \email{carlos.pellicer@ect.ufrn.br}

\affiliation{$^{1}$Departamento de Física Teórica and IFIC, Centro Mixto Universidad de Valencia--CSIC. Universidad
de Valencia, Burjassot--46100, Valencia, Spain}
\affiliation{$^{2}$Departamento de Física, Universidade Federal da Paraíba, Caixa Postal 5008, 58051-970, João Pessoa, Paraíba, Brazil}
\affiliation{$^{3}$Physics Department, State University of Tetovo, 
Ilinden Street nn, 1200, Tetovo, North Macedonia}
\affiliation{$^{4}$Engineering School of Lorena - University of Sao Paulo (EEL-USP), 
Estrada Municipal do Campinho Nº 100, Campinho, 
CEP: 12602-810, Lorena, SP - Brazil}
\affiliation{$^{5}$Departamento de Matem\'{a}ticas, Universidad Cat\'{o}lica del Norte, Avenida Angamos 0610, Casilla 1280, Antofagasta, Chile}
\affiliation{$^{6}$Institute of Systems Science, Durban University of Technology, PO Box 1334,  
Durban 4000, South Africa}
\affiliation{$^{7}$ Institute for Theoretical Physics
and Cosmology, Zhejiang University of Technology, Hangzhou 310023,
China.}
\affiliation{$^{8}$ Department of Mathematics, COMSATS University Islamabad, Lahore-Campus, Lahore-54000, Pakistan}
\affiliation{$^{9}$ Escola de Ciência e Tecnologia, Universidade Federal do Rio Grande do Norte,
Campus Universitário Lagoa Nova, CEP 59078-970, Natal, RN, Brazil}


\begin{abstract}
This study derives a novel family of charged black hole solutions featuring short-- and long-range modifications. These ones are achieved through a Yukawa--like gravitational potential modification and a nonsingular electric potential incorporation. The short--range corrections encode quantum gravity effects, while the long-range adjustments simulate gravitational effects akin to those attributed to dark matter. Our investigation reveals that the total mass of the black hole undergoes corrections owing to the apparent presence of dark matter mass and the self--adjusted electric charge mass. Two distinct solutions are discussed: a regular black hole solution characterizing small black holes, where quantum effects play a crucial role, and a second solution portraying large black holes at considerable distances, where the significance of Yukawa corrections comes into play. Notably, these long-range corrections contribute to an increase in the total mass and hold particular interest as they can emulate the role of dark matter. Finally, we explore the phenomenological aspects of the black hole. Specifically, we examine the influence of electric charge and Yukawa parameters on thermodynamic quantities, the quasinormal modes for the charged scalar perturbations as well as for the vector perturbations, analysis of the geodesics of light/massive particles, and the accretion of matter onto the charged black hole solution.

\end{abstract}
\maketitle


\section{Introduction}

Black holes are one of the most interesting predictions of General Relativity, a fundamental gravitational theory. Their existence was largely confirmed by the successive detection of gravitational waves coming from binary mergers by the LIGO~\cite{LIGO} and Virgo collaborations as well as the observation of the shadow of supermassive black holes such as M87 and Sgr A* by the Event Horizon Telescope (EHT) array~\cite{EventHorizonTelescope:2019dse,  EventHorizonTelescope:2020qrl,  EventHorizonTelescope:2021srq,  EventHorizonTelescope:2022wkp}. The gravitational signal coming from a binary merger has a final ringdown phase described by the resulting object's quasinormal modes (QNM). These modes are characterized by a complex frequency,  $\omega=\omega_{\Re}-I \omega_{\Im}$, where the real part represents the oscillation of the wave, and the imaginary part amounts to its damping rate. Remarkably, these frequencies depend only on the parameters that identify the black hole. Furthermore, they can be used to test the dynamical stability of a black hole by considering different types of perturbations and studying the wave response of the perturbed geometry. When $\omega_{\Im} >0$, the stability of the object is guaranteed. 

On astronomical scales, photons emitted from a luminous source near a black hole (BH) can spiral towards the event horizon or be deflected away. Critical geodesic trajectories are identified as unstable spherical orbits that delineate the boundary between these two outcomes. By observing these pivotal paths of photons, we can capture the image of a BH's shadow \cite{Takahashi:2004xh,  Hioki:2009na,  Brito:2015oca,  Cunha:2015yba,  Ohgami:2015nra,  Moffat:2015kva,  Abdujabbarov:2016hnw,  Cunha:2018acu,  Mizuno:2018lxz,  Tsukamoto:2017fxq,  Psaltis:2018xkc,  Amir:2018pcu,  Gralla:2019xty,  Bambi:2019tjh,  Cunha:2019ikd,  Khodadi:2020jij,  Perlick:2021aok,  Vagnozzi:2022moj,  Saurabh:2020zqg,  Jusufi:2020cpn,  Tsupko:2019pzg} like those brought out by the EHT Collaboration. Moreover, a connection between shadows and quasinormal modes at the eikonal limit has been proposed and analytically tested~\cite{Jusufi:2019ltj,Cuadros-Melgar:2020kqn}.

On the other hand,  from cosmological observations, it has been discovered that cold dark matter is invisible and can only be detected through its gravitational effects. One might naturally then ask about the impact of dark matter on black holes. Despite decades of research, the nature and origin of this mysterious substance remain unclear \cite{Gonzalez:2023rsd,  Bond:1984fp,  Trimble:1987ee}. Despite persistent endeavors, directly detecting particles of dark matter has proven challenging with its presence primarily inferred from gravitational effects on galaxies and other expansive cosmic structures. Conversely, another intriguing element, dark energy, is introduced to account for the observed accelerated expansion of the universe \cite{Carroll:1991mt}. This proposal gains robust support from an abundance of observational evidence \cite{SupernovaCosmologyProject:1997zqe,  SupernovaSearchTeam:1998fmf,  SupernovaCosmologyProject:1998vns}. Mysteries persist in fundamental physics, especially regarding dark matter and energy. Scalar fields' role in depicting the universe has gained attention in inflationary scenarios \cite{Guth:1980zm}. Expanding upon the $\Lambda$CDM framework, a generalized model incorporating a quintessence scalar field has been introduced to probe these cosmic enigmas \cite{Ratra:1987rm,  Parsons:1995kt,  Rubano:2001xi,  Saridakis:2008fy,  Cai:2009zp,  WaliHossain:2014usl,  Barrow:2016qkh}. Additionally, multi-scalar field models have emerged as versatile tools capable of shedding light on various cosmic epochs \cite{Elizalde:2004mq,  Elizalde:2008yf,  Skugoreva:2014ena,  Saridakis:2016mjd,  Paliathanasis:2019luv,  Banerjee:2022ynv,  Santos:2023eqp}. Furthermore, other efforts have been made to formulate a unified description encompassing matter-dominated and dark energy-driven epochs, leading to the development of scalar-torsion theories with a Hamiltonian foundation \cite{Leon:2022oyy}. 

One alternative view in the literature proposes modifying Einstein's equations to develop new theories of gravity. These theories can provide explanations for observed phenomena and challenge the traditional framework \cite{CANTATA:2021ktz,  Leon:2009rc,  DeFelice:2010aj,  Clifton:2011jh,  Capozziello:2011et,  DeFelice:2011bh,  Xu:2012jf,  Bamba:2012cp,  
Leon:2012mt,  Kofinas:2014aka,  Bahamonde:2015zma,  Momeni:2015uwx,  Cai:2015emx,  Krssak:2018ywd,  Dehghani:2023yph}. Dark matter has been used to explain the peculiar flatness observed in galaxy rotation curves \cite{Salucci:2018eie}, for example. Modified Newtonian Dynamics (MOND) \cite{Milgrom:1983ca} was one of the initial theories proposed for this phenomenon, modifying Newton's classical law of gravitation \cite{Ferreira:2009eg,  Milgrom:2003,  Tiret:2007kq,  Kroupa:2010hf,  Cardone:2010ru,  Richtler:2011zk}. Moreover, there are additional intriguing proposals within the realm of dark matter. These encompass concepts such as superfluid dark matter \cite{Berezhiani:2015bqa} and the captivating notion of a Bose–Einstein condensate \cite{Boehmer:2007um}, among other innovative ideas.

Recently, Yukawa potential was used in the context of cosmology \cite{Jusufi:2023xoa,  Gonzalez:2023rsd}; here we shall point out some of the most important results, which are the main physical motivation for studying the Yukawa modified potential in the context of BHs. In particular, using the Yukawa potential, it was argued that one can obtain effectively the $\Lambda$CDM model in cosmology. This can be achieved since there exists an expression that relates baryonic matter, effective dark matter, and dark energy as \cite{Jusufi:2023xoa,  Gonzalez:2023rsd},  
\begin{equation}\label{DMCC}
   \Omega_{DM}(z)= \sqrt{2\, \Omega_{B,  0}  \Omega_{\Lambda,  0}}{(1+z)^3},  
\end{equation}
where it was introduced the definition for the dark energy density (here we restore $c$ for a moment) \cite{Jusufi:2023xoa,  Gonzalez:2023rsd},  
\begin{equation}
    \Omega_{\Lambda,  0}= \frac{c^2}{\lambda^2 H^2_0}\frac{\alpha}{(1+\alpha)^2}\,  .
\end{equation}
One can interpret these results in the following way: dark matter arises due to the modified Newton law quantified by $\alpha$ and $\Omega_B$. If we have no contribution at all from the baryonic matter, i.e., $\Omega_{B,  0}=0$, automatically, the effect of dark matter also vanishes, i.e., $\Omega_{DM}=0$. This implies that dark matter can be viewed as an apparent effect. The Yukawa cosmology was confronted with observational data in \cite{Gonzalez:2023rsd}, where, among other things, it was shown that one can get $\Omega_{\Lambda,  0}\simeq 0.69$ and $\Omega_{DM}\simeq 0.26$, precisely the $\Lambda$CDM parameters from cosmological observations.  Motivated by these ideas,  in \cite{Filho:2023abd} the Yukawa potential was used to study the implications of the BH geometry. The present work aims to find other exciting consequences by combining the gravitational singular/nonsingular Yukawa potential and a singular/nonsingular electric potential. We strive to study charged BH solutions with Yukawa corrections and investigate their thermodynamic properties and dynamical stability by calculating quasinormal modes. We also explore the relation between quasinormal frequencies, shadows of the charged metric, and matter accretion onto the BH.

The paper is organized as follows. In Section \ref{sectII}, we obtain two classes of charged BH solutions with Yukawa potential. Further, Section \ref{sectIII} uses the Kerr-Schild method to verify the BH solution in the region $r\gg \ell_0$. In Section \ref{sectIV} we study the thermodynamics of these solutions. In Sections \ref{sectV}, \ref{sectVI}, \ref{sectVII}, and \ref{sectVIII} we explore the role of electric charge on quasinormal modes, geodesics, BH shadows, and accretion of matter onto the BH, respectively. Finally, in Section \ref{sectIX} we discuss our findings. 


\section{Charged black hole solution with Yukawa potential}
\label{sectII}
In the present paper we shall consider a potential modified due to the mass of the graviton, namely, the modified Yukawa-type potential \cite{Jusufi:2023xoa},
\begin{equation}
\Phi(r)=-\frac{G M }{\sqrt{r^2+\ell_0^2}}\left(1+\alpha\,  e^{-\frac{r}{\lambda}}\right).
\end{equation}

Notice that the wavelength of the massive graviton is represented by $\lambda=\hbar/(m_g c)$ and $\ell_0$ is a deformed parameter of Planck length size. Let us see how the spacetime geometry around the 
BH is modified in this theory. The general \textit{Ansatz} in the case of a static, spherically symmetric source reads,  
\begin{equation}
\mathrm{d}s^2
= -f(r) \mathrm{d}t^2 +\frac{ \mathrm{d}r^2}{f(r)} +r^2(\mathrm{d}\theta^2+\sin^2\theta \mathrm{d}\phi^2). \label{metric}
\end{equation}

The energy density of the modified matter can be computed from 
$
\rho(r)=\frac{1}{4\pi} \Delta \Phi(r)$, so that we obtain the following relation \cite{Filho:2023abd},
\begin{eqnarray}
    \rho(r)=\frac{e^{-\frac{r}{\lambda}}M \alpha \mathcal{X}}{4 \pi r \lambda^2 (r^2+\ell_0^2)^{5/2}}+ \frac{3M \ell_0^2}{4 \pi (r^2+\ell_0^2)^{5/2} }
    \label{energydensity},  
\end{eqnarray}
where we defined $\mathcal{X}=(2\lambda-r)\ell_0^4+(3 \lambda^2 r+2\lambda r^2-2r^3)\ell_0^2-r^5$. The energy density, therefore, consists of two terms: the first one is proportional to $\alpha$ and it is important in large distances; the second one is proportional to $\ell_0$ and plays an important role in short distances instead. In particular, if we neglect the long-range modification, i.e., as a special case of our results, only the second term remains consistent with \cite{Filho:2023abd}. 

    \subsection{Singular BH solution: BH solution in the spacetime region: $\ell_0\ll  r \leq \lambda$}
In order to find a BH solution in this region, let us expand only the first factor in Eq. \eqref{energydensity} in a series around $\ell_0$. After taking into account the leading order terms in $\alpha$ and $\ell_0$, we get
\begin{eqnarray}
    \rho(r)=-\frac{M \alpha}{4 \pi r \lambda^2} e^{-\frac{r}{\lambda}}+\frac{3M \ell_0^2}{4 \pi (r^2+\ell_0^2)^{5/2} }+\mathcal{O}(\ell_0^2 \alpha).
\end{eqnarray}

Moreover, it is worth mentioning that the negative sign reflects that the energy conditions are violated inside the BH. On the other hand, we assume that the Einstein field equations with a cosmological constant still hold, 
\begin{eqnarray}
    G_{\mu \nu}+\Lambda g_{\mu \nu}=8\pi (T_{\mu \nu}+T_{\mu \nu}^{em})
\end{eqnarray}
so that the effects of the effective dark matter are encoded in the stress-energy tensor. To generate a new solution that includes the electric charge, we need to take into account the energy contribution of the electric field encoded in the electromagnetic potential given by,
 \begin{eqnarray}
  \Phi^{em}(r)=-\frac{Q}{\sqrt{r^2+\ell_0^2}}(1+\alpha_{em} e^{-\frac{r}{\lambda}}).
\end{eqnarray}
For the photon we consider $m_{ph}=0$, yielding $\lambda_{ph} \to \infty$. The potential is thus simplified as
\begin{eqnarray}
  \Phi^{em}(r)=-\frac{\mathcal{Q}}{\sqrt{r^2+\ell_0^2}}
\end{eqnarray}
where $\mathcal{Q}=Q (1+\alpha_{em})$. 
Considering the spherical symmetry in the spacetime metric \eqref{metric}, we will assume that the form of the stress-energy tensor of the electromagnetic field has the form,
\begin{equation}\label{Tem}
 T_{\mu \nu}^{em}=\frac{1}{4 \pi}\left(F_{\mu \sigma}{F_{\nu}}^{\sigma}    -\frac{1}{4}g_{\mu \nu}F_{\rho \sigma}F^{\rho \sigma}\right)
\end{equation}
Thus, the energy density of the stress-energy field along with the pressure components are \cite{Gaete:2022ukm},
\begin{eqnarray}
    {\rho}^{em}(r)&=&-{p_r}^{em}(r)={p_t}^{em}(r)=\frac{\mathcal{Q}^2 r^2}{8 \pi (r^2+\ell_0^2)^3}.
\end{eqnarray}

One can check that the solution 
\begin{eqnarray}\notag
f(r)&=&1-\frac{2M r^2}{(r^2+\ell_0^2)^{3/2}}+\frac{\mathcal{Q}^2 r^2 \mathcal{F}(r)}{(r^2+\ell_0^2)^2}\\
&-&\frac{2M \alpha (r+\lambda) e^{-\frac{r}{\lambda}}}{r \lambda} -\frac{\Lambda r^2}{3}, \label{eq13}
\end{eqnarray}
where $M$ is a mass parameter and 
\begin{equation}
 \mathcal{F}(r)=\frac{5}{8}+\frac{3 \ell_0^2}{8 r^2}-\frac{3 (r^2+\ell_0^2)^2 }{8 \ell_0 r^3}\arctan\left(\frac{r}{\ell_0}\right),  
\end{equation}
solves the gravitational field equations
    \begin{eqnarray}
        \frac{r f'(r)+f(r)-1}{r^2}&+&\Lambda-\frac{2 M \alpha}{ r \lambda^2} e^{-\frac{r}{\lambda}}\\\notag
        &+&\frac{\mathcal{Q}^2 r^2}{(r^2+\ell_0^2)^3}
        +\frac{6M \ell_0^2}{ (r^2+\ell_0^2)^{5/2} }=0, \label{eq.(5)}
    \end{eqnarray}
Since $\ell_0$ is of Planck length order, i.e., $\ell_0\sim 10^{-35}\,  m$,  in large distances and astrophysical BHs with large mass $M$, we must have $r\gg\ell_0$; then, we can neglect the effect of $\ell_0$. In the present work we will focus precisely on astrophysical BHs. On the other hand, $\lambda$ is of Mpc order and, therefore, it is important on large distances. Thus, we can perform a series expansion around $\varepsilon=1/\lambda$, yielding the result that the solution turns out to be 
 \begin{equation}
       f(r)\simeq 1-\frac{2M(1+\alpha)+\frac{3\pi Q^2}{16 \ell_0}}{r}+\frac{\mathcal{Q}^2}{r^2}+\frac{\alpha M  r}{\lambda^2}-\frac{\Lambda r^2}{3} \label{eq.(7)}
    \end{equation}
The apparent dark matter mass is the extra term in mass $M \alpha$. This mass is not real, but it appears only due to the modification of the gravitational potential. It is, therefore, natural to define the true or  physical mass of the BH to be 
\begin{eqnarray}
\mathcal{M}=M(1+\alpha)+\frac{3\pi \mathcal{Q}^2}{32 \ell_0}. \label{eq17}
\end{eqnarray}
As we can see, the mass contains a new term proportional to
the regularized self-energy of the electrostatic field as was shown in \cite{Gaete:2022ukm}, but also the apparent dark matter term as was recently shown in \cite{Filho:2023abd}.
Hence, for an observer located far away from the BH with mass $\mathcal{M}$, we have the metric 
\begin{equation}
\label{metric16}
       f(r)\simeq 1-\frac{2\mathcal{M}}{r}+\frac{\mathcal{Q}^2}{r^2}+\frac{ \mathcal{M}_{\alpha} r}{(1+\alpha) \lambda^2}-\frac{\Lambda r^2}{3},  
    \end{equation}
where
\begin{eqnarray}
    \mathcal{M}_{\alpha}=\alpha \left( \mathcal{M}-\frac{3\pi \mathcal{Q}^2}{32 \ell_0} \right).
\end{eqnarray}
We can express the general metric solution \eqref{eq13} in terms of  the physical mass as follows,
\begin{eqnarray}\notag
f(r)&=&1-\frac{2\mathcal{M}r^2}{(1+\alpha)(r^2+\ell_0^2)^{3/2}}+\frac{\mathcal{Q}^2 r^2\, \mathcal{H}(r)}{(r^2+\ell_0^2)^2}\\
&-& \frac{2\mathcal{M}_{\alpha} (r+\lambda)\, e^{-\frac{r}{\lambda}}}{r \lambda (1+\alpha)}-\frac{\Lambda r^2}{3},  
\end{eqnarray}
valid in the region $\ell_0\ll  r\leq \lambda$, where
\begin{eqnarray}\notag
 \mathcal{H}(r)&=&\frac{5}{8}+\frac{3 \ell_0^2}{8 r^2}+\frac{3 \pi  (r^2+\ell_0^2)^{1/2} }{16 (1+\alpha) \ell_0}\\
 &-&\frac{3 (r^2+\ell_0^2)^2 }{8 \ell_0 r^3}\arctan\left(\frac{r}{\ell_0}\right).
\end{eqnarray}
We should notice that this solution is singular in this region. In the particular case $\alpha=0$ we obtain the charged BH without a cosmological constant in T-duality recently reported by Gaete-Jusufi-Nicolini \cite{Gaete:2022ukm} and with the presence of the cosmological constant outlined in \cite{Jusufi:2022ukt}. 

\subsection{Nonsingular BH solution: BH solution in the spacetime region: $\ell_0\leq r\ll  \lambda$}
In this case, the metric is important and describes small black holes (for such black holes, quantum effects are expected to play an important role) in the region $ \ell_0 \leq r \ll  \lambda$, when viewed from outside. In particular, we can set $e^{r/\lambda} \to 1$, then, the potential turns out to be,
\begin{equation}
\Phi(r)=-\frac{G M }{\sqrt{r^2+\ell_0^2}}\left(1+\alpha\right).
\end{equation}
Then, we obtain for the energy density, 
\begin{eqnarray}
    \rho(r)=\frac{3 M (1+\alpha) \ell_0^2}{4 \pi (r^2+\ell_0^2)^{5/2} }.
\end{eqnarray}
And the spacetime geometry becomes, 
\begin{eqnarray}\notag
f(r)&=&1-\frac{2M(1+\alpha)r^2}{(r^2+\ell_0^2)^{3/2}}+\frac{\mathcal{Q}^2 r^2\, \mathcal{F}(r)}{(r^2+\ell_0^2)^2}-\frac{\Lambda r^2}{3}.
\end{eqnarray}
To obtain the BH mass, we can rewrite it by expanding in series around $\ell_0$, namely,
 \begin{equation}
       f(r)\simeq 1-\frac{2M(1+\alpha)+\frac{3\pi Q^2}{16 \ell_0}}{r}+\frac{\mathcal{Q}^2}{r^2}-\frac{\Lambda r^2}{3} \label{eq24}
    \end{equation}
If we again use the mass definition \eqref{eq17}, we can finally write the metric in this case as follows,
\begin{eqnarray}
f(r)&=&1-\frac{2\mathcal{M}r^2}{(r^2+\ell_0^2)^{3/2}}+\frac{\mathcal{Q}^2 r^2\, \mathcal{H}(r)}{(r^2+\ell_0^2)^2}-\frac{\Lambda r^2}{3}, \label{eq25}
\end{eqnarray}
which is valid in the region $ \ell_0 \leq r \ll  \lambda$, where 
\begin{eqnarray}\notag
 \mathcal{H}(r)&=&\frac{5}{8}+\frac{3 \ell_0^2}{8 r^2}+\frac{3 \pi  (r^2+\ell_0^2)^{1/2} }{16\ell_0}\\
 &-&\frac{3 (r^2+\ell_0^2)^2 }{8 \ell_0 r^3}\arctan\left(\frac{r}{\ell_0}\right).
\end{eqnarray}
This solution is regular at the center which we can verify by calculating the following curvature scalars,
\begin{eqnarray}
    \lim_{r \to 0}R_{ab}g^{ab}=\frac{24 \mathcal{M}}{\ell_0^3}+4\Lambda-\frac{9 \pi \mathcal{Q}^2}{4 \ell_0^4},  
\end{eqnarray}
\begin{eqnarray}\notag
    \lim_{r \to 0}R_{abcd}R^{abcd}&=&\frac{96 \mathcal{M}^2}{\ell_0^6}-\frac{18 \mathcal{M} \mathcal{Q}^2 \pi}{\ell_0^7}+\frac{27 \pi^2 \mathcal{Q}^4}{32 \ell_0^8}\\
    &+&\frac{8 \Lambda^2}{3}-\frac{3 \Lambda \pi \mathcal{Q}^2}{\ell_0^4}+\frac{32 \Lambda \mathcal{M}}{\ell_0^3}.
\end{eqnarray}

Again, in the special case $\alpha=0$ this result is in perfect agreement with the charged BH solution reported in \cite{Gaete:2022ukm} and with the presence of the cosmological constant reported in \cite{Jusufi:2022ukt}. Also, in the limit $\ell_0 \to 0$ and $\mathcal{H}\to 1$ we get the Ayon-Beato and Garcia metric \cite{Ayon-Beato:1998hmi}, whose  approximated form reads, 
 \begin{equation}
       f(r)\simeq 1-\frac{2 \mathcal{M}}{r}+\frac{\mathcal{Q}^2}{r^2}-\frac{\Lambda r^2}{3}
    \end{equation}
which is the well-known RN-dS metric. Metric \eqref{eq24} describes the spacetime geometry around the BH up to some critical distance $r=r_c$ when the metric deviates due to the long-range effect of $\alpha/\lambda$ term.

In the next section we will obtain the singular solution describing large black holes using another interesting method, the Kerr-Schild \textit{ Ansatz}.

\section{Charged solution using the Kerr-Schild method}
\label{sectIII}
The singular solution shown in the previous section can also be obtained through other methods. One of them, the Kerr-Schild \textit{Ansatz}, will be applied here to obtain this new solution. This \textit{Ansatz} appeared for the first time when obtaining the Kerr metric from a flat spacetime and was extended to other cases by the same authors, Kerr and Schild~\cite{KSansatz}. Subsequently, Taub made an important generalization incorporating a general background metric $g_{\mu\nu}$~\cite{TAUB1981326}. Thus, to produce new solutions, it is necessary to add to this background metric a term proportional to a null geodesic vector $\ell_\mu$ and a scalar function $H$, 
\begin{equation}\label{ksnewg}
\tilde g_{\mu\nu}=g_{\mu\nu} + 2 H(r) \ell_\mu \ell_\nu.     
\end{equation}
Although initially this method is perturbative, the new metric $\tilde g_{\mu\nu}$ is, in fact, an exact solution of the field equations. The advantage of the Generalized Kerr-Schild (GKS) method derives from the linearity of the new Einstein equations when written in covariant-contravariant form. Nevertheless, we should mention that this method has also been applied successfully to non-linear gravity, where the resulting equations are quadratic in $H$~\cite{PhysRevD.81.126010}. 

As a first step, let us determine the null geodesic vector through the following conditions,  
\begin{equation}\label{nullgeo}
\ell_{\mu;\nu} \ell^\nu = 0,   \qquad \ell_\mu \ell^\mu = 0.  
\end{equation}
Assuming only a radial dependence, namely, $\ell_\mu (r)$, Eqs. \eqref{nullgeo} yield,  
\begin{equation}
\ell_\mu = \left( E, \sqrt{\frac{E^2}{f^2(r)}-\frac{K^2}{r^2 f(r)}},  K,  0\right),  
\end{equation}
where $E$ and $K$ are constants (as we want to obtain a diagonal metric in the end, we will make $K=0$ in what follows) and $f(r)$ is the background metric given by~\cite{Filho:2023abd},  
\begin{equation}\label{bgmet}
f(r)=1-\frac{2M (1+\alpha)}{r}+\frac{M \alpha r}{\lambda^2}-\frac{\Lambda r^2}{3},   
\end{equation}
which obeys the Einstein equations plus a cosmological constant sourced by a stress-energy tensor from a Yukawa-type potential in the regime $r \gg \ell_0$, where $\lambda$ has a substantial contribution. 

Now, we can calculate the non-null covariant-contravariant components of the Einstein tensor using the new metric \eqref{ksnewg},  
\begin{eqnarray}
G_t ^t = G_r ^r &=& -\frac{1}{r^2} + \frac{f(r)}{r^2} + \frac{f'(r)}{r} - \frac{2E^2 H(r)}{r^2} \nonumber \\ 
&&- \frac{2E^2 H'(r)}{r},   \label{Gtt}  \\
G_\theta ^\theta = G_\phi ^\phi &=& \frac{f'(r)}{r} + \frac{f''(r)}{2} - \frac{2E^2 H'(r)}{r} \nonumber \\
&&- E^2 H''(r).  \label{Gthth}
\end{eqnarray}

In order to generate a new solution endowed with charge, we add an electromagnetic component to the stress-energy tensor in the form \eqref{Tem}, so that,  
\begin{equation}
T_\mu ^{\nu\,  (em)} = \frac{1}{4\pi}\begin{pmatrix}
\frac{-Q^2}{2r^4} & 0 & 0 & 0 \\
0 & \frac{-Q^2}{2r^4} & 0 & 0 \\
0 & 0 & \frac{Q^2}{2r^4} & 0 \\
0 & 0 & 0 & \frac{Q^2}{2r^4}
\end{pmatrix}\,  .
\end{equation}
Thus, the new Einstein equations in covariant-contravariant form reduce to only two equations, namely,  
\begin{eqnarray}
\frac{f'(r)}{r} - \frac{2H'(r)E^2}{r} + \frac{f(r)}{r^2} - \frac{2E^2 H(r)}{r^2} - \frac{1}{r^2} && \nonumber \\
+ \Lambda - \frac{2M\alpha}{r\lambda^2} + \frac{Q^2}{r^4} = 0,  && \\
-H''(r) E^2 - \frac{2H'(r)E^2}{r} + \frac{f''(r)}{2} + \frac{f'(r)}{r} && \nonumber \\
+ \Lambda - \frac{M\alpha}{r\lambda^2} - \frac{Q^2}{r^4} = 0. &&
\end{eqnarray}
Solving for $H(r)$ we obtain,  
\begin{equation}
H(r) = -\frac{Q^2}{2E^2 r^2} + \frac{c}{r},  
\end{equation}
being $c$ an integration constant. Thus, the resulting new metric $\tilde g_{\mu\nu}$ becomes,  
\begin{eqnarray}
\tilde{\mathrm{d}s}^2 &=& \left( -f(r) -\frac{Q^2}{r^2} + \frac{2cE^2}{r} \right) \mathrm{d}t^2 \nonumber \\
&&+ \left( -\frac{2Q^2}{r^2 f(r)} + \frac{4cE^2}{r f(r)}\right) \mathrm{d}t\,  \mathrm{d}r \nonumber \\
&&+ \left( \frac{1}{f(r)} - \frac{Q^2}{r^2 f^2(r)} + \frac{2cE^2}{r f^2(r)} \right) dr^2 \nonumber \\
&&+r^2(d\theta^2+\sin^2\theta \mathrm{d}\phi^2).
\end{eqnarray}
Finally, we perform the transformation $\mathrm{d}t = \mathrm{d}\tilde t + u(r) \mathrm{d}r$ with
\begin{equation}
u(r) = -\frac{2cE^2r - Q^2}{f(r)(2cE^2 r -f(r) r^2 -Q^2)},    
\end{equation}
and make $cE^2=\frac{3\pi Q^2}{32 \ell_0}$  to obtain 
\begin{equation}
\tilde{\mathrm{d}s}^2 = -\tilde f(r) \mathrm{d}\tilde t^2 + \frac{\mathrm{d}r^2}{\tilde f(r)} +r^2(\mathrm{d}\theta^2+\sin^2\theta \mathrm{d}\phi^2),  
\end{equation}
with 
\begin{equation}
\tilde f(r) = 1-\frac{2M(1+\alpha)+\frac{3\pi Q^2}{16 \ell_0}}{r} + \frac{Q^2}{r^2} + \frac{\alpha Mr}{\lambda^2} - \frac{\Lambda r^2}{3},  
\end{equation}
which is the same as Eq. \eqref{eq.(7)}. The choice we took to fix these constants comes from the regularized self-energy of the electric field and, as we saw in the previous section, it is responsible for the BH mass modification.

Once we have well established the charged solutions corrected by a Yukawa-type potential, we will continue our study analyzing the thermodynamic aspects brought by the new geometry in the next section.

\section{Thermodynamics}
\label{sectIV}
\subsection{Thermodynamics of the regular BH solution}
In this section our focus shifts to investigating the thermodynamics of the regular solution \eqref{eq25}. Following the approach in \cite{Gaete:2022ukm}, we introduce a function $\mathcal{G}(r)$ for this purpose as follows,
\begin{equation}
f(r)=1-\frac{2\mathcal{M}}{r}\ \mathcal{G}(r),  
\label{eq:genmetric}
\end{equation}
where $\mathcal{G}(r)$ is given by, 
\begin{equation}
\mathcal{G}(r)=\frac{r^3}{\left(r^2+\ell_0^2\right)^{\frac{3}{2}}}\left[1 - \frac{\mathcal{Q}^2 \mathcal{H}(r)}{2 \mathcal{M} \left(r^2+\ell_0^2\right)^{\frac{1}{2}}}+\frac{\Lambda\left(r^2+\ell_0^2\right)^{\frac{3}{2}}}{6 \mathcal{M}}\right].
\end{equation}
Essentially, in the limit of large $r$ one can consistently reproduce the RN-dS case, where $\mathcal{G} \to 1 - \mathcal{Q}^2/(2\mathcal{M}r)+\Lambda r^3/(6\mathcal{M})$. 

Now, we can calculate the Hawking temperature as,
\begin{equation}
T=\frac{1}{4\pi r_+}\left[1-r_+\  \frac{\mathrm{d}\mathcal{G}(r_+)/\mathrm{d}r_+}{\mathcal{G}(r_+)}  \right].
\end{equation}

\begin{figure}
    \centering
    \includegraphics[scale=0.55]{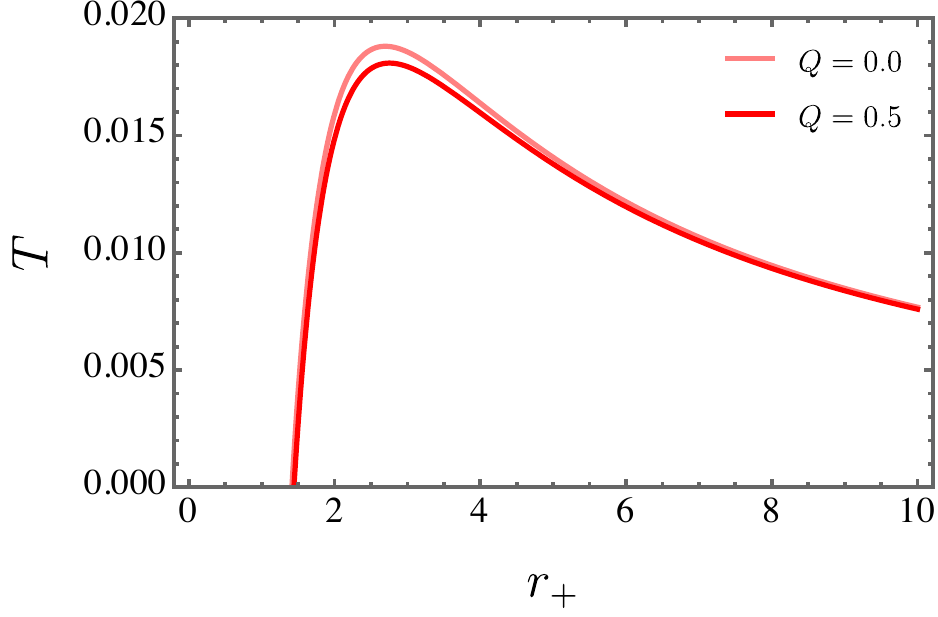}
    \caption{\textit{Hawking} temperature as a function of $r_{+}$ for the regular black  hole. We have set $\mathcal{M}=1$ and $\ell_0=1$ along with $\Lambda = 10^{-5}$.}\label{hawking_regular}
\end{figure}

\begin{figure}
    \centering
     \includegraphics[scale=0.55]{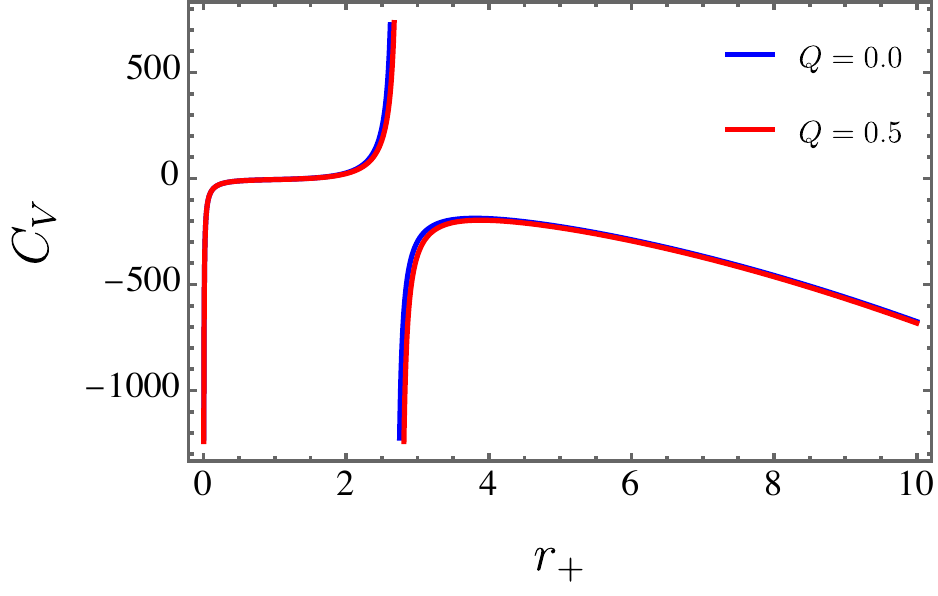}
    \caption{Heat capacity as a function of $r_{+}$ for the regular black  hole. We have set $\mathcal{M}=1$ and $\ell_0=1$ along with $\Lambda = 10^{-5}$.}\label{heat_label}
\end{figure}

One can also write the entropy of the BH as a function of $\mathcal{G}$ as follows,
\begin{equation}
S(r_+)=2\pi \int_{r_\mathrm{ext}}^{r_+} \frac{r\mathrm{d}r}{\mathcal{G}(r)},  
\end{equation}
which is consistent with the area law up to certain short-scale deformations. In addition, it is possible to calculate the heat capacity with the general formula being,
\begin{equation}
C_V=\left.\frac{\partial \mathcal{M}}{\partial T}\right|_{r_+}.
\end{equation}
We can utilize \eqref{eq:genmetric} to derive $\mathcal{M}=r_+/2\mathcal{G}(r)$ from the horizon equation. Subsequently, one can express
\begin{equation}
C_V(r_+)=\frac{2\pi r_+}{\mathcal{G}(r_+)}\ T \left(\frac{\mathrm{d}T}{\mathrm{d}r_+}\right)^{-1}.
\label{eq:heatcap}
\end{equation}
The \textit{Hawking} temperature and the heat capacity are depicted in Figs. \ref{hawking_regular} and \ref{heat_label}, respectively. 
From Eq.(\ref{eq:heatcap}) can be demonstrated that the system undergoes a phase transition at the maximum temperature indicated by $\mathrm{d}T/\mathrm{d}r_+=0$, as we can also see in both figures. Moreover, this phase transition is a second-order one. In addition, the BH is stable when the heat capacity is positive and unstable when it is negative. Besides, it is noteworthy that the presence of charge slightly shifts the location of the phase transition (i.e., vertical asymptote) to larger values of $r_+$.

\subsection{Thermodynamics of the singular BH solution}

We begin by calculating the horizons to provide the thermodynamic analysis of Eq. \eqref{metric16}. Four horizons appear after considering $f(r)=0$. Nevertheless, if the condition of $0<\mathcal{Q}<1$ is taken into account, we have three physical horizons, i.e., the Cauchy horizon ($\Tilde{r}_{+}$), the event horizon ($r_{+}$), and the cosmological horizon ($\overset{\nsim}{r}_{+}$), which are written as follows,

\ie
\begin{split}
& \Tilde{r}_{+} =  \frac{3 \mathcal{M}_{\alpha }}{4 (\alpha +1) \lambda ^2 \Lambda } - \frac{1}{2} \left(\frac{\eta }{\gamma  \left(-\alpha  \lambda ^2 \Lambda -\lambda ^2 \Lambda \right)} \right. \\
& \left. + \frac{\gamma }{3 \sqrt[3]{2} \left(-\alpha  \lambda ^2 \Lambda -\lambda ^2 \Lambda \right)}\right)^{1/2} + \frac{1}{2} \left(   \frac{\eta }{\gamma  \left(-\alpha  \lambda ^2 \Lambda -\lambda ^2 \Lambda \right)} \right. \\
& \left. - \frac{\gamma }{3 \sqrt[3]{2} \left(-\alpha  \lambda ^2 \Lambda -\lambda ^2 \Lambda \right)}  - \frac{27 \mathcal{M}_{\alpha }^3}{(\alpha +1)^3 \lambda ^6 \Lambda ^3}-\frac{36 \mathcal{M}_{\alpha }}{(\alpha +1) \lambda ^2 \Lambda ^2} \right. \\
& \left. + \frac{48 \mathcal{M}}{\Lambda }         \right)^{1/2} \times \left( \frac{1}{4\sqrt{ \left(\frac{\eta }{\gamma  \left(-\alpha  \lambda ^2 \Lambda -\lambda ^2 \Lambda \right)}+\frac{\gamma }{3 \sqrt[3]{2} \left(-\alpha  \lambda ^2 \Lambda -\lambda ^2 \Lambda \right)}\right.}} \right)^{1/2},  
\end{split}
\fe

\ie
\begin{split}
& r_{+} =  \frac{3 \mathcal{M}_{\alpha }}{4 (\alpha +1) \lambda ^2 \Lambda } + \frac{1}{2} \left(\frac{\eta }{\gamma  \left(-\alpha  \lambda ^2 \Lambda -\lambda ^2 \Lambda \right)} \right. \\
& \left. + \frac{\gamma }{3 \sqrt[3]{2} \left(-\alpha  \lambda ^2 \Lambda -\lambda ^2 \Lambda \right)}\right)^{1/2} - \frac{1}{2} \left(   \frac{\eta }{\gamma  \left(-\alpha  \lambda ^2 \Lambda -\lambda ^2 \Lambda \right)} \right. \\
& \left. - \frac{\gamma }{3 \sqrt[3]{2} \left(-\alpha  \lambda ^2 \Lambda -\lambda ^2 \Lambda \right)}  + \frac{27 \mathcal{M}_{\alpha }^3}{(\alpha +1)^3 \lambda ^6 \Lambda ^3}+\frac{36 \mathcal{M}_{\alpha }}{(\alpha +1) \lambda ^2 \Lambda ^2} \right. \\
& \left. - \frac{48 \mathcal{M}}{\Lambda }         \right)^{1/2} \times \left( \frac{1}{4\sqrt{ \left(\frac{\eta }{\gamma  \left(-\alpha  \lambda ^2 \Lambda -\lambda ^2 \Lambda \right)}+\frac{\gamma }{3 \sqrt[3]{2} \left(-\alpha  \lambda ^2 \Lambda -\lambda ^2 \Lambda \right)}\right.}} \right)^{1/2},  
\end{split}
\fe

\ie
\begin{split}
& \overset{\nsim}{r}_{+} =  \frac{3 \mathcal{M}_{\alpha }}{4 (\alpha +1) \lambda ^2 \Lambda } + \frac{1}{2} \left(\frac{\eta }{\gamma  \left(-\alpha  \lambda ^2 \Lambda -\lambda ^2 \Lambda \right)} \right. \\
& \left. + \frac{\gamma }{3 \sqrt[3]{2} \left(-\alpha  \lambda ^2 \Lambda -\lambda ^2 \Lambda \right)}\right)^{1/2} + \frac{1}{2} \left(   \frac{\eta }{\gamma  \left(-\alpha  \lambda ^2 \Lambda -\lambda ^2 \Lambda \right)} \right. \\
& \left. - \frac{\gamma }{3 \sqrt[3]{2} \left(-\alpha  \lambda ^2 \Lambda -\lambda ^2 \Lambda \right)}  + \frac{27 \mathcal{M}_{\alpha }^3}{(\alpha +1)^3 \lambda ^6 \Lambda ^3}+\frac{36 \mathcal{M}_{\alpha }}{(\alpha +1) \lambda ^2 \Lambda ^2} \right. \\
& \left. - \frac{48 \mathcal{M}}{\Lambda }         \right)^{1/2} \times \left( \frac{1}{4\sqrt{ \left(\frac{\eta }{\gamma  \left(-\alpha  \lambda ^2 \Lambda -\lambda ^2 \Lambda \right)}+\frac{\gamma }{3 \sqrt[3]{2} \left(-\alpha  \lambda ^2 \Lambda -\lambda ^2 \Lambda \right)}\right.}} \right)^{1/2},  
\end{split}
\fe
being 
\ie
\begin{split}
\nonumber
\eta = &\frac{3}{\Lambda }+\frac{\alpha  \lambda ^2+\lambda ^2}{-\alpha  \lambda ^2 \Lambda -\lambda ^2 \Lambda }+\frac{9 \mathcal{M}_{\alpha }^2}{4 (\alpha +1)^2 \lambda ^4 \Lambda ^2} \\
& + 3 \sqrt[3]{2} \left(\alpha ^2 \lambda ^4+2 \alpha  \lambda ^4+\lambda ^4-4 \alpha ^2 \lambda ^4 \Lambda  \mathcal{Q}-8 \alpha  \lambda ^4 \Lambda  \mathcal{Q} \right. \\
& \left. -4 \lambda ^4 \Lambda  \mathcal{Q}+6 \lambda ^2 \mathcal{M} \mathcal{M}_{\alpha }+6 \alpha  \lambda ^2 \mathcal{M} \mathcal{M}_{\alpha }\right),  
\end{split}
\fe
and
\ie
\begin{split}
\gamma= & \left( 54 \left(\alpha  \lambda ^2+\lambda ^2\right)^3+972 \left(-\alpha  \lambda ^2 \Lambda -\lambda ^2 \Lambda \right) \left(\alpha  \lambda ^2 \mathcal{M}+\lambda ^2 \mathcal{M}\right)^2  \right. \\
& \left. -648 \left(\alpha  \lambda ^2+\lambda ^2\right) \left(-\alpha  \lambda ^2 \Lambda -\lambda ^2 \Lambda \right) \left(\alpha  \lambda ^2 \mathcal{Q}+\lambda ^2 \mathcal{Q}\right) \right. \\
& \left.  + 486 \left(\alpha  \lambda ^2+\lambda ^2\right) \mathcal{M}_{\alpha } \left(\alpha  \lambda ^2 \mathcal{M}+\lambda ^2 \mathcal{M}\right) \right. \\
& \left. +729 \mathcal{M}_{\alpha }^2 \left(\alpha  \lambda ^2 \mathcal{Q}+\lambda ^2 \mathcal{Q}\right)     \left(  -4 \left( 9 \left(\alpha  \lambda ^2+\lambda ^2\right)^2  \right.\right.\right.  \\
& \left.\left.\left. +36 \left(-\alpha  \lambda ^2 \Lambda -\lambda ^2 \Lambda \right) \left(\alpha  \lambda ^2 \mathcal{Q}+\lambda ^2 \mathcal{Q}\right)    \right.\right.\right.  \\
& \left.\left.\left. +54 \mathcal{M}_{\alpha } \left(\alpha  \lambda ^2 \mathcal{M}+\lambda ^2 \mathcal{M}\right)\right)^{3}          + \left(  54 \left(\alpha  \lambda ^2+\lambda ^2\right)^3  \right.\right.\right.  \\
& \left.\left.\left. +972 \left(-\alpha  \lambda ^2 \Lambda -\lambda ^2 \Lambda \right) \left(\alpha  \lambda ^2 \mathcal{M}+\lambda ^2 \mathcal{M}\right)^2  \right.\right.\right.  \\
& \left.\left.\left.  -648 \left(\alpha  \lambda ^2+\lambda ^2\right) \left(-\alpha  \lambda ^2 \Lambda -\lambda ^2 \Lambda \right) \left(\alpha  \lambda ^2 \mathcal{Q}+\lambda ^2 \mathcal{Q}\right)  \right.\right.\right.  \\
& \left.\left.\left.  + 486 \left(\alpha  \lambda ^2+\lambda ^2\right) \mathcal{M}_{\alpha } \left(\alpha  \lambda ^2 \mathcal{M}+\lambda ^2 \mathcal{M}\right)   \right.\right.\right.  \\
& \left.\left.\left.  +729 \mathcal{M}_{\alpha }^2 \left(\alpha  \lambda ^2 \mathcal{Q}+\lambda ^2 \mathcal{Q}\right)        \right)^{2}         \right)^{1/2}      \right)^{1/3}  .
\end{split}
\fe

Nevertheless, we shall regard only the event horizon to perform the calculation. In this way, the thermodynamic properties can accurately be calculated as follows.


\subsubsection{The Hawking temperature}

\begin{figure}
    \centering
    \includegraphics[scale=0.45]{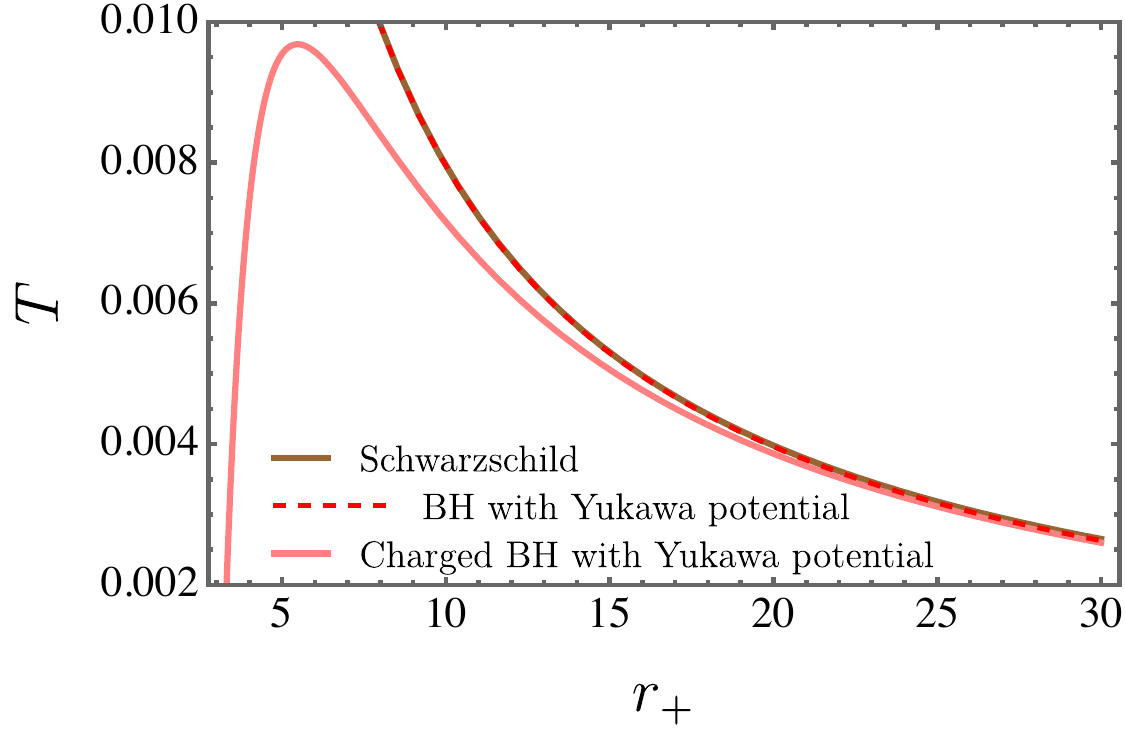}
    \includegraphics[scale=0.45]{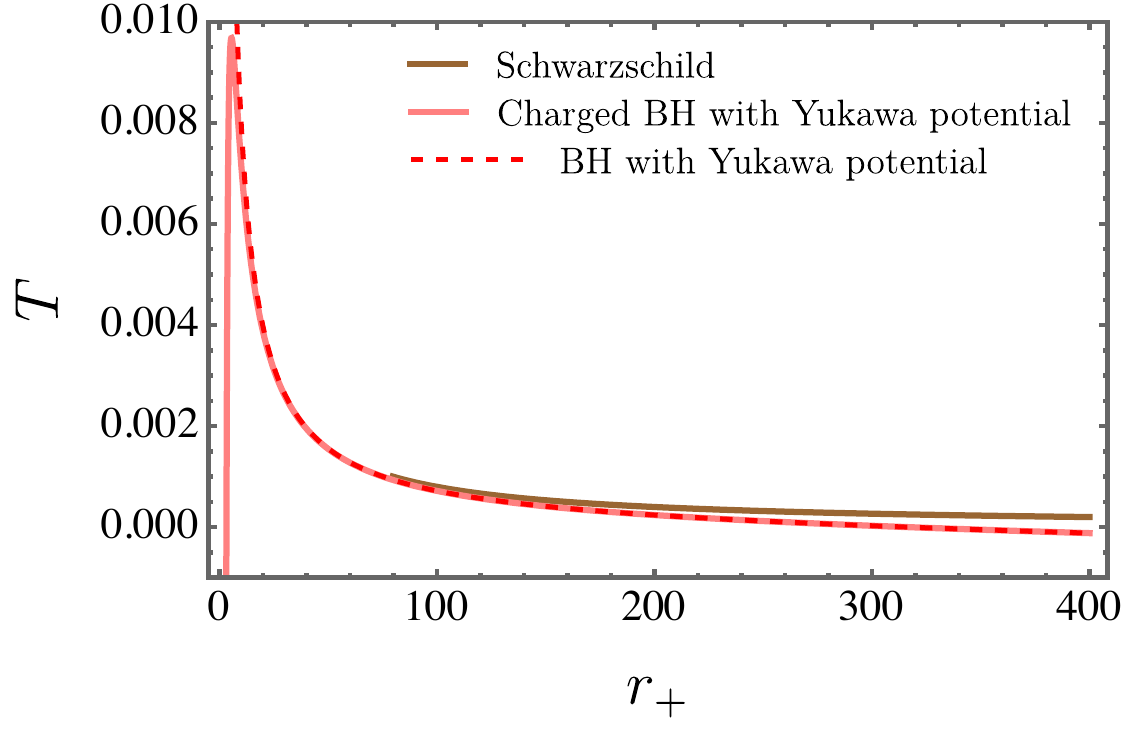}
    \caption{The \textit{Hawking} temperature for the charged BH with Yukawa potential in comparison with the Schwarzschild case and a Yukawa BH when different ranges of $r_{+}$ are taken into account. Here, it is regarded $\alpha =1, \lambda =10^5, \Lambda = 10^{-5}, \mathcal{Q}=10$, and $\mathcal{M}_{\alpha} =1$.}
    \label{htemp}
\end{figure}

As defined in Eq. \eqref{metric16}, the metric exhibits a timelike Killing vector represented as $\xi = \partial/\partial t$. This characteristic imparts the metric with a conserved quantity intrinsically associated with the Killing vector denoted by $\xi$ \cite{heidari2023gravitational,sedaghatnia2023thermodynamical}. The construction of such a conserved quantity is achievable by harnessing this Killing vector, 
\ie
\label{killing}
\nabla^\nu(\xi^\mu\xi_\mu) = -2\kappa\xi^\nu.
\fe
Within this context, it is important to notice that the covariant derivative is denoted by \(\nabla_\nu\) and \(\kappa\) remains constant along the orbits defined by \(\xi\). More explicitly, this implies that the Lie derivative of \(\kappa\) with respect to \(\xi\) is invariably zero,
\ie
\label{liederivative}
\mathcal{L}_\xi\kappa = 0.
\fe
Significantly, the surface gravity, denoted by $\kappa$, remains unchanged across the horizon. In the coordinate basis, the timelike Killing vector components are succinctly expressed as $\xi^{\mu} = (1, 0, 0, 0)$. The representation of the surface gravity for the metric prescribed in Eq. \eqref{metric16} unfolds as follows,
\ie 
\label{surfacegravity}
\kappa = {\left.\frac{f^{\prime}(r)}{2} \right|_{r = {r^{(2)}_{+}}}}.
\fe 
Next, knowing that the \textit{Hawking} temperature is written as $T = \kappa/2\pi$, we can straightforwardly derive
\ie
\label{hawkingsurface}
T = \frac{1}{4 \pi } \left( \frac{2 r_{+} \mathcal{M} - 2 \mathcal{Q}}{r_{+}^3} - \frac{2 \Lambda  r_{+}}{3}+\frac{\mathcal{M}_{\alpha }}{(\alpha +1) \lambda ^2}\right),  
\fe
where, if we consider $f(r_{+})=0$, $\mathcal{M}$ can be expressed as
\ie
\begin{split}
\mathcal{M} = & \frac{1}{96 l_{0} r_{+} \left(-2 \alpha  \lambda ^2-2 \lambda ^2+\alpha  r_{+}^2\right)} \\ 
& \times  \left\{ 
 -96 \lambda ^2 l_{0} \mathcal{Q}-96 \alpha  \lambda ^2 l_{0} r_{+}^2-96 \lambda ^2 l_{0} r_{+}^2+9 \pi  \alpha  \mathcal{Q}^2 r_{+}^3   \right.  \\
&\left. -96 \alpha  \lambda ^2 l_{0} \mathcal{Q}+32 \alpha  \lambda ^2 \Lambda  l_{0} r_{+}^4+32 \lambda ^2 \Lambda  l_{0} r_{+}^4 \right\}.
\end{split}
\fe
In Fig. \ref{htemp} the \textit{Hawking} temperature of a charged BH with a Yukawa potential is analyzed in comparison to both the Schwarzschild case and a Yukawa BH, considering various ranges of the event horizon. The values used for the parameters were $\alpha =1, \lambda =10^5, \Lambda = 10^{-5}, \mathcal{Q}=10$, and $\mathcal{M}_{\alpha} =1$.

\subsubsection{Entropy}

Now, let us calculate the \textit{Hawking} temperature via the first law of thermodynamics,
\ie
\begin{split}
\label{tf}
\Tilde{T}_{f} & = \frac{\mathrm{d} \mathcal{M}}{\mathrm{d} \mathcal{S}} = \frac{1}{2\pi r_{+}} \frac{\mathrm{d}\mathcal{M}}{\mathrm{d}r_{+}} \\
& = \frac{1}{48 \pi  l_{0} r_{+}^3 \left(\alpha  r_{+}^2-2 (\alpha +1) \lambda ^2\right)^2} \\ 
& \times \left\{  (\alpha +1) \lambda ^2 \left[ - 9 \pi  \alpha  \mathcal{Q}^2 r_{+}^3 + 8 l_{0}  \left( 3 \alpha  r_{+}^2 \left(3 \mathcal{Q}+r_{+}^2\right) \right.\right.\right.\\
& \left.\left.\left. + \alpha  \Lambda  r_{+}^6-6 (\alpha +1) \lambda ^2 \Lambda  r_{+}^4 + 6 (\alpha +1) \lambda ^2 \left(r_{+}^2-\mathcal{Q}\right)\right)     \right]    \right\}.
\end{split}
\fe
Nevertheless, the determined temperature contradicts the value presented in Eq.\eqref{hawkingsurface}. Employing the methodology elucidated in Ref. \cite{Ma:2014qma} and assuming the validity of the area law, we apply the corrected first law of thermodynamics to establish the temperature for regular BHs \cite{araujo2023analysis,filho2023implications}. The subsequent formulation encapsulates the corrected temperature as prescribed by the following first law,
\ie
\Upsilon(r_{+},  \mathcal{M},  \mathcal{Q},  \alpha,  \lambda,  \Lambda) \mathrm{d}\mathcal{M} = \Tilde{{T}}_{f}\, \mathrm{d}S.
\fe
Here, $\Tilde{{T}}_{f}$ signifies the corrected version of the \textit{Hawking} temperature derived from the first law of thermodynamics while $S$ represents the entropy. As outlined in Ref. \cite{Ma:2014qma}, the general formula for $\Upsilon(r_{+},  \mathcal{M},  \alpha,  \lambda,  \Lambda)$ is expressed as follows,
\begin{equation}
\label{entropycorrection}
\Upsilon(r_{+},  \mathcal{M},  \mathcal{Q},  \alpha,  \lambda,  \Lambda) = 1 + 4\pi \int^{\infty}_{r_{+}} r^{2} \frac{\partial T^{0}_{0}}{\partial  \mathcal{M}} \mathrm{d}r.
\end{equation}

In this context the notation $T^{0}_{0}$ pertains to the stress-energy component related to the energy density. Consequently, Eq. \eqref{entropycorrection} can be explicitly evaluated as
\ie
\begin{split}
\Upsilon(r_{+},  \mathcal{M},  \mathcal{Q},  \alpha,  \lambda,  \Lambda) =  \frac{12 l_{0} r_{+}^3 \left(\alpha  r_{+}^2-2 (\alpha +1) \lambda ^2\right)^2 \Tilde{\epsilon}}{(\alpha +1) \lambda ^2 \Tilde{\Phi}},  
\end{split}
\fe
where
\ie
\begin{split}
\Tilde{\epsilon} & = \frac{\alpha  \left(\mathcal{M}-\frac{3 \pi  \mathcal{Q}^2}{32 l_{0}}\right)}{(\alpha +1) \lambda ^2}+\frac{2 r_{+} \mathcal{M}-2 \mathcal{Q}}{r_{+}^3}-\frac{2 \Lambda  r_{+}}{3},  
\end{split}
\fe
and
\ie
\begin{split}
\Tilde{\Phi} & =  8 l_{0} \left(\alpha  \Lambda  r_{+}^6-6 (\alpha +1) \lambda ^2 \Lambda  r_{+}^4+6 (\alpha +1) \lambda ^2 \left(r_{+}^2-\mathcal{Q}\right) \right. \\
& \left. + 3 \alpha  r_{+}^2 \left(3 \mathcal{Q}+r_{+}^2\right)\right)-9 \pi  \alpha  \mathcal{Q}^2 r_{+}^3.
\end{split}
\fe
Categorically, now, the \textit{Hawking} temperatures Eq. \eqref{hawkingsurface} and Eq. \eqref{tf} can be in agreement with each other by considering,
\begin{equation}
T = \Upsilon(r_{+},  \mathcal{M},  \mathcal{Q},  \alpha,  \lambda,  \Lambda) \Tilde{T}_{f}\,,
\end{equation}
in such a way that, 
\begin{equation}
S = \int \frac{\Upsilon(r_{+},  \mathcal{M},  \alpha,  \lambda,  \Lambda)}{\overset{\nsim}{T}_{f}} \mathrm{d} \mathcal{M} = \pi r_{+}^{2} = \frac{A}{4},  
\end{equation}
which recovers the usual \textit{Bekenstein-Hawking} area law.


\subsubsection{The heat capacity}

Actually, in the pursuit of investigating additional thermodynamic quantities, particular attention is drawn to the heat capacity \cite{furtado2023thermal,araujo2023thermodynamical}, which can be articulated as
\ie
\begin{split}
& C_{V} = T \frac{\mathrm{d}S}{\mathrm{d}T} \\
& = \frac{1}{(\alpha +1) \lambda ^2 \left(-3 \mathcal{Q}+\Lambda  r_{+}^4+r_{+}^2\right)} \\  
& \times \left[  4 \pi  r_{+}^2 \left((\alpha +1) \lambda ^2 \mathcal{Q}+(\alpha +1) \lambda ^2 \Lambda  r_{+}^4-2 r_{+}^3 \mathcal{M}_{\alpha} \right. \right. \\
& \left. \left. - (\alpha +1) \lambda ^2 r_{+}^2\right)    \right].
\end{split}
\fe

\begin{figure}
    \centering
    \includegraphics[scale=0.45]{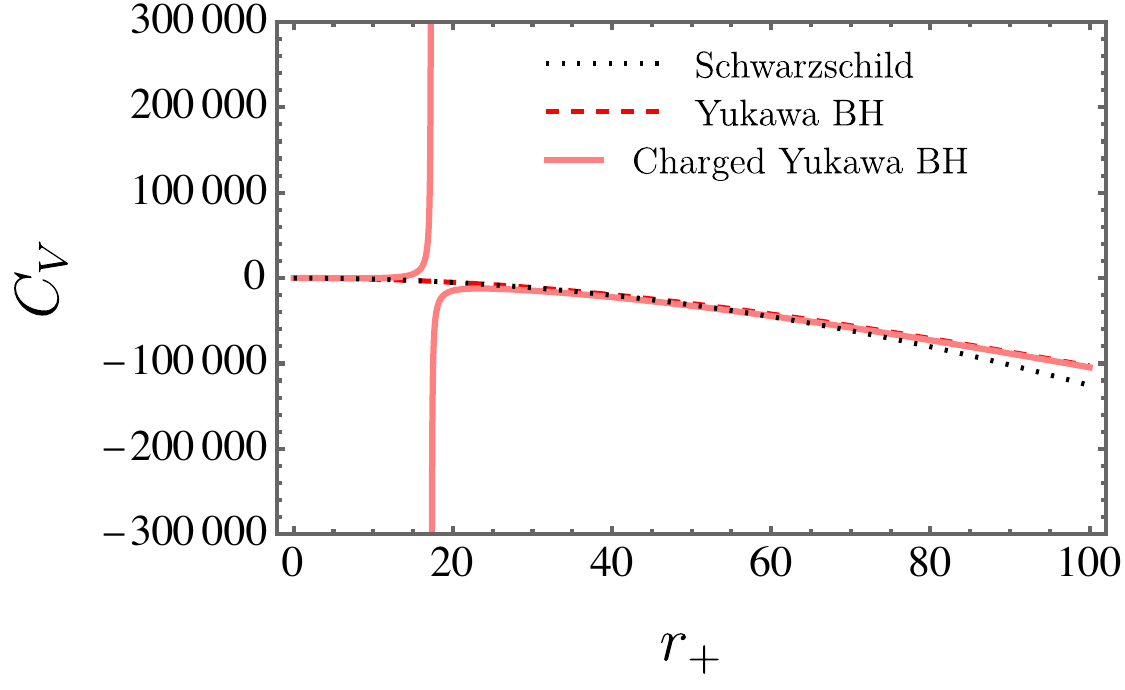}
    \caption{Heat capacity as a function of $r_{+}$ for $\lambda = 10^{5}$, $\mathcal{Q}=10$, $\Lambda = 10^{-5}$, $\alpha = 1$, and $\mathcal{M}_{\alpha}=1$}
    \label{he1}
\end{figure}

\begin{figure}
    \centering
    \includegraphics[scale=0.45]{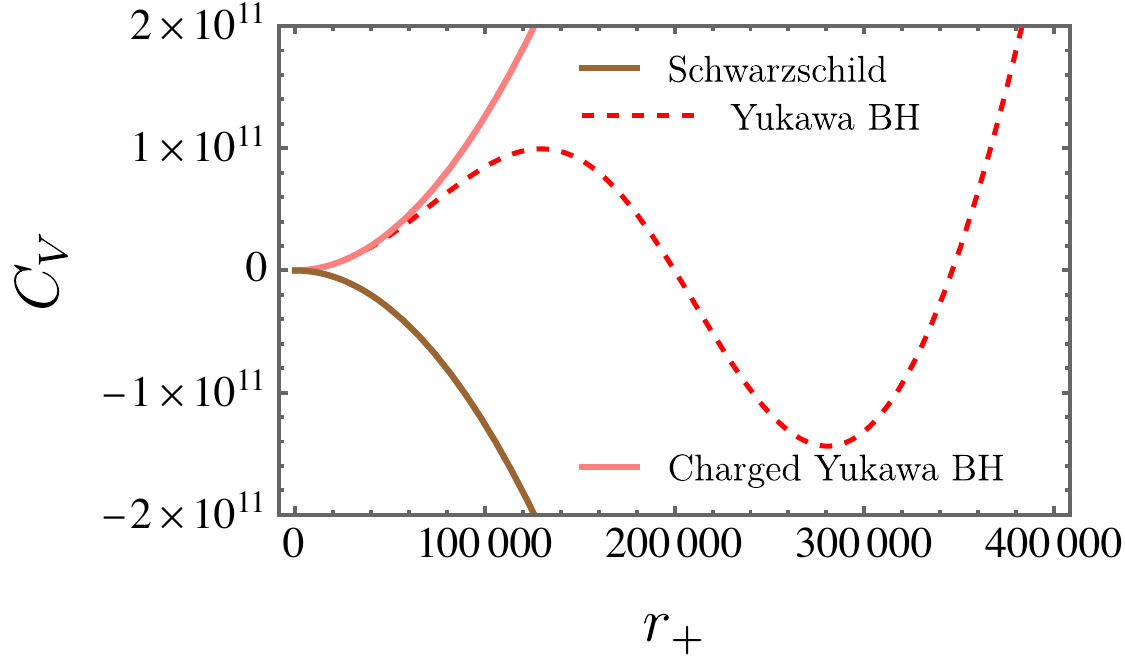}
    \caption{Heat capacity as a function of $r_{+}$ (huge values) for $\lambda = 10^{5}$, $\mathcal{Q}=10$, $\Lambda = 10^{-5}$, $\alpha = 1$, and $\mathcal{M}_{\alpha}=1$}
    \label{he2}
\end{figure}

To visualize our outcome, we present Figs. \ref{he1} and \ref{he2}. Also, we compare our new results with those presented in the literature, i.e., the Schwarzschild and the Yukawa black hole recently proposed. Specifically, Fig. \ref{he1} shows a second-order phase transition. Notice that no first-order transitions differ from the uncharged Yukawa black hole case.

Having analyzed the thermodynamic stability, we will consider the dynamical stability of our solutions in the next section by studying their quasinormal modes' response when perturbed by external fields. 

\section{Quasinormal modes}
\label{sectV}

In the aftermath of the ringdown phase an intriguing phenomenon emerges, denoted as \textit{quasinormal} modes. These modes unveil unique oscillation patterns that persist unaltered by the initial perturbations. They serve as a manifestation of the inherent characteristics of the system, stemming from the innate oscillations of spacetime, free from the influence of particular initial conditions.

Diverging from \textit{normal} modes tied to closed systems, \textit{quasinormal} modes find their association with open systems. Consequently, these modes undergo a gradual energy dissipation process via the emission of gravitational waves. Their mathematical representation involves conceptualizing them as poles within the complex Green function in the AdS/CFT conjecture context.

When establishing their frequencies, the quest involves seeking solutions to the wave equation within a system characterized by a background metric $g_{\mu\nu}$. Nevertheless, obtaining analytical solutions for these modes typically poses a formidable challenge \cite{heidari2023exploring,hassanabadi2023gravitational,reis2023exploring}.

Various techniques have been proposed in the scientific literature for obtaining solutions. The WKB (Wentzel-Kramers-Brillouin) approach is particularly prominent among these methods. Its evolution traces back to the pioneering work of Will and Iyer \cite{Iyer:1986np,  Iyer:1986nq} with subsequent advancements up to the sixth order accomplished by Konoplya \cite{Konoplya:2003ii} and to the thirteen order by Matyjasek and Opala~\cite{Matyjasek:2017psv}. In what follows, we will discuss charged scalar and electromagnetic perturbations in the background of large black holes.


\subsection{Charged scalar perturbations}

This section will closely follow the discourse presented in \cite{Cuadros-Melgar:2021sjy}. Our focus will be on analyzing a perturbation of a charged scalar field denoted as $\Psi$, which can be described by the Klein-Gordon equation,
\begin{equation}\label{kge}
[(\nabla^\nu - iqA^\nu)(\nabla_\nu -iqA_\nu) -\mu^2] \Psi =0.
\end{equation}
Here the parameters $\mu$ and $q$ represent the mass and charge of the field $\Psi$, respectively, while the electromagnetic potential $A_\mu = (-\mathcal{Q}/\sqrt{r^2+\ell_0^2}, 0, 0, 0)$ characterizes the BH. 
Throughout our discussion we will employ the background geometry described by Eq. \eqref{metric16}, resulting in a spherically symmetric metric. The perturbed scalar field is denoted by $\Psi(t, r, \theta, \phi) = \frac{1}{r} \psi(r,t) e^{im\phi} S(\theta)$, where $S(\theta) = P_\ell^m(\cos\theta)$ represents the associated Legendre polynomials. Consequently, the time-radial component of Eq. \eqref{kge} takes the form,
\begin{eqnarray}
\label{feq1}
&& - \frac{\partial^2 \psi}{\partial t^2} + 2 i q \Phi(r) \frac{\partial \psi}{\partial t} + \frac{1}{f(r)} \frac{\partial}{\partial r} \left( \frac{1}{f(r)} \frac{\partial \psi}{\partial r} \right) \\\notag
&-& \left[ - q^2 \Phi^2 +\mu^2 f + \ell (\ell
+1)\frac{f}{r^2} + \frac{f^\prime}{r}\right] \psi = 0. 
\end{eqnarray}
Furthermore, in terms of the tortoise coordinate defined as $dr_* = dr/f(r),$
Eq. \eqref{feq1} becomes,
\begin{equation}\label{psieq}
- \frac{\partial^2 \psi}{\partial t^2} + 2 i q \Phi(r) \frac{\partial \psi}{\partial t} +
\frac{\partial^2 \psi}{\partial r_*^2} - \mathcal{V}(r) \, \psi = 0,  
\end{equation}
where the potential can be written as 
\begin{equation}\label{pot}
\mathcal{V}(r) = - q^2 \Phi^2(r) + f(r)
\left[ \frac{\ell (\ell +1)}{r^2} + \mu^2 + \frac{f'(r)
    }{r}  \right]. 
\end{equation}

\subsubsection{Superradiance condition}
In order to get some information on the possibility of superradiance due to the presence of a charged scalar field interacting with the charged Yukawa BH, we will further simplify Eq.(\ref{psieq}) choosing the time dependence of the field as $\psi(r, t) = e^{-i\omega t} X(r)$. 
Thus,  in terms of the tortoise coordinate the radial part of Eq. \eqref{kge} becomes,
\begin{equation}\label{Xeq}
\frac{d^2 X}{dr_*^2} + \tilde{\mathcal{V}} \, X = 0,  
\end{equation}
where the potential can be written as \cite{Cuadros-Melgar:2021sjy},
\begin{equation}\label{pot}
\tilde{\mathcal{V}} = \left( \omega - \frac{q \mathcal{Q}}{r} \right)^2 - f(r)
\left[ \frac{\ell (\ell +1)}{r^2} + \mu^2 + \frac{f'(r)
    }{r}  \right]. 
\end{equation}
The behavior of this potential at the event and cosmological horizons can lead us to a superradiance condition. Thus, in these limits we have~\cite{Cuadros-Melgar:2021sjy},
\begin{eqnarray}
\tilde{\mathcal{V}} & \rightarrow &  (\omega
+q\Phi_h)^2\,, \quad r \rightarrow r_+ \\
\tilde{\mathcal{V}} & \rightarrow & (\omega + q\Phi_c)^2\,, \quad r \rightarrow r_c\,,
\end{eqnarray}
where $\Phi_h$ and $\Phi_c$ are the electromagnetic potentials at the event and cosmological horizons, respectively. Since there are only ingoing waves at the event horizon, we can write the near-horizon solutions for the scalar field as,
\begin{eqnarray}
X(r_*) =  \left\{
\begin{array}{ll} 
{\cal T} e^{-i (\omega+q\Phi_h) r_*} \quad \hbox{ as }
r\rightarrow r_h\\
{\cal R} e^{i(\omega+q\Phi_c)r_*} + {\cal I} e^{-i
  (\omega+q\Phi_c)r_*} \quad \hbox{ as } r\rightarrow r_c 
\end{array} \right.\,,
\end{eqnarray}
where ${\cal I}$, ${\cal R}$, and ${\cal T}$ are the
incident, reflected, and transmitted wave amplitudes,
respectively. Now, due to the conservation of the Wronskian, $W=X
\partial_{r_*} X^* - X^* \partial_{r_*} X$, we can obtain a relation among these amplitudes, which reads,
\begin{equation}
|{\cal R}|^2 = |{\cal I}|^2 - \frac{\kappa}{\lambda} |{\cal T}|^2 \,,
\end{equation}
being $\kappa=\omega + q\Phi_h$ and $\lambda = \omega+q\Phi_c$. Then, if we are looking for superradiance, the condition  $\kappa/\lambda<0$ must be fulfilled, so
that
\begin{equation}\label{src}
\frac{qQ}{r_c} < \omega < \frac{qQ}{r_h} \,,
\end{equation}
which is the same condition found in the RN-dS and the Kiselev black holes, noticing that the position of the cosmological horizon depends on the parameters of each model. 
It is important to stress that the frequency in Eq.(\ref{src}) is real. Moreover, if there are quasinormal modes, solutions of Eq.(\ref{psieq}), whose real frequency falls inside the range given by Eq.(\ref{src}), these modes are called superradiant modes. Also, the corresponding imaginary part of the  quasinormal frequency will dictate if this mode is stable or not as we will see in the next subsection. 

\subsubsection{Numerical results}
In order to calculate the quasinormal modes from Eq. \eqref{psieq} we employ 
the finite difference method, setting
$\psi(t, r_*) = \psi(t_0 + j \Delta t, r_{*0} + k \Delta r_*) = \psi_{j,k}$
we approximate the derivatives as finite differences
and write Eq. \eqref{psieq} as,
\begin{eqnarray} \notag
&-& \frac{\psi_{j+1,k} - 2 \psi_{j,k} + \psi_{j-1,k}}{\Delta t^2}
+ 2 i q \Phi_k \frac{\psi_{j+1,k} - \psi_{j-1,k}}{2 \Delta t} \\ 
&+& \frac{\psi_{j,k+1} - 2 \psi_{j,k} + \psi_{j,k-1}}{\Delta r_*^2}
- \mathcal{V}_k \psi_{j,k} = 0 \ , \label{fdm}
\end{eqnarray}
where $\Phi_k = \Phi(r(r_{*0} + k \Delta r_*))$ and  $\mathcal{V}_k = \mathcal{V}(r(r_{*0} + k \Delta r_*))$ are obtained by the inverse of the tortoise coordinate.
The initial conditions set the value of $\psi_{j,k}$ for $j=0$ and $j=1$. For
$j>1$ we use Eq. \eqref{fdm} since we know the values of $\psi_{j,k}$
at the two previous points of $j$. 

\begin{figure}
    \centering
    \includegraphics[scale=0.45]{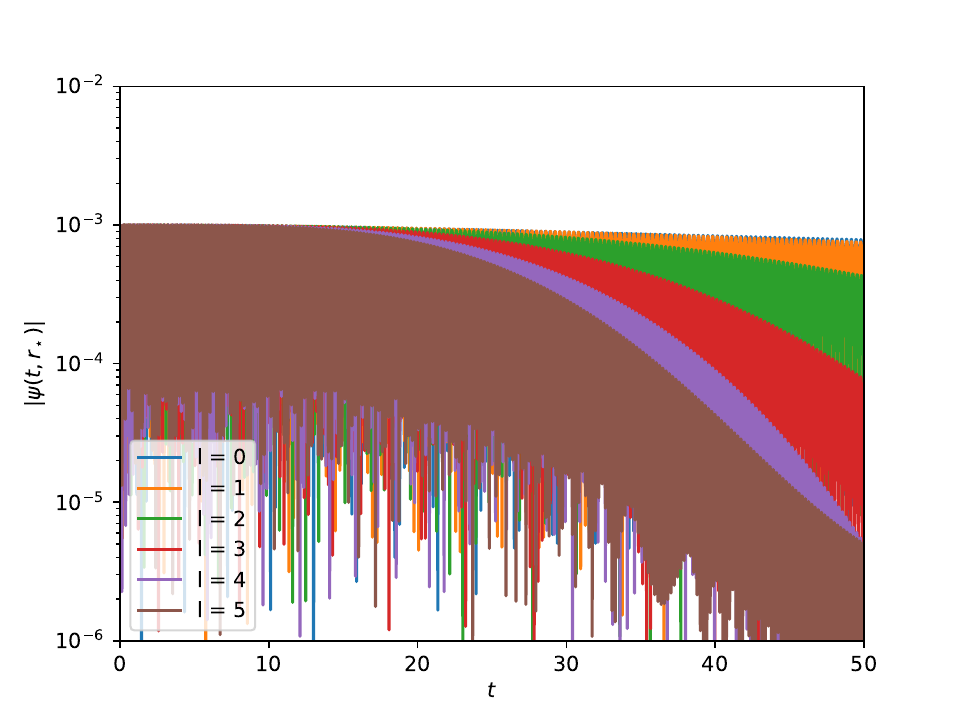}
    \caption{Scalar perturbation as a function of $t$ for $\lambda = 10^{5}$, $\mathcal{Q}=0.1$, $\Lambda = 10^{-5}$, $\alpha = 1$, $\mathcal{M}_{\alpha}=1$, $\mu = 0.1$ and $q = 0.1$.}
    \label{figscpertl}
\end{figure}

\begin{figure}
    \centering
    \includegraphics[scale=0.45]{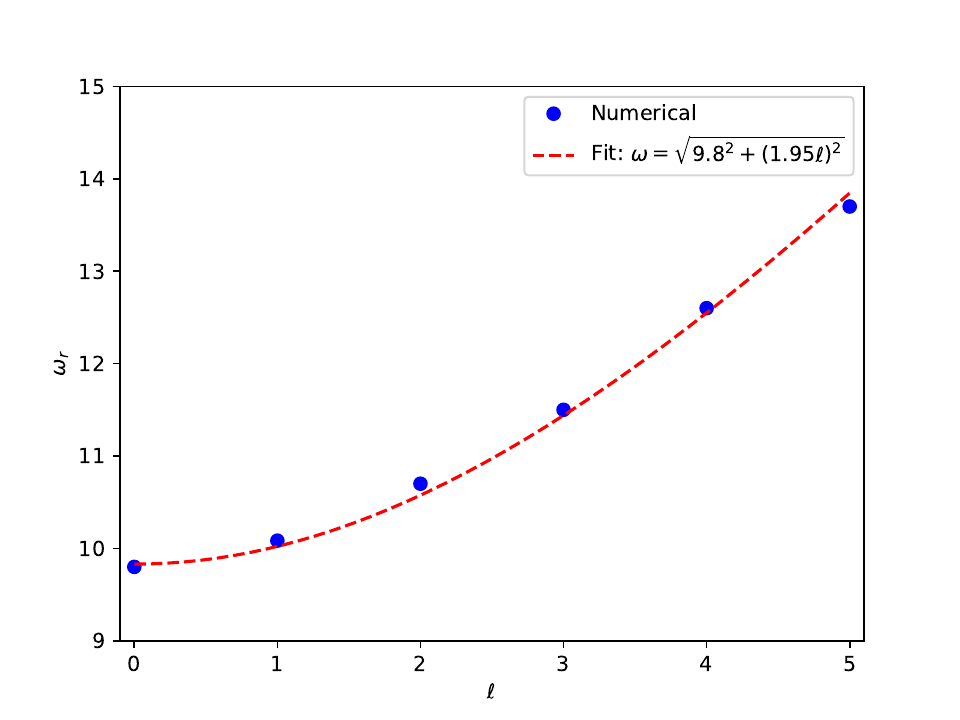}
    \caption{Real part of the frequencies shown in Fig. \ref{figscpertl} }
    \label{figfreqrl}
\end{figure}

Choosing ${\cal M}_\alpha = 1$, $Q = 0.1$, $\alpha = 1$, $\ell_0 = 1$, $\lambda = 10^{5}$, $\Lambda = 10^{-5}$, $\mu = 0.1$ and $q = 0.1$ we get the behavior shown in Fig.
\ref{figscpertl}, where we see that the quasinormal modes decay very slowly. 
The real part of the frequencies 
is displayed in Fig. \ref{figfreqrl}, it shows a behavior that can be approximated by an arc of a hyperbola. The imaginary part of the frequencies is not constant, \textit{i. e.}, the amplitude does not behave like an exponential decay, so we cannot employ the usual regression method to calculate these modes. However, as our main goal is to determine the stability of the model, it is enough to know that all the modes considered here are decaying. Moreover, we see
that modes with higher $\ell$ decay faster as one should expect.

\begin{figure}
    \centering
    \includegraphics[scale=0.45]{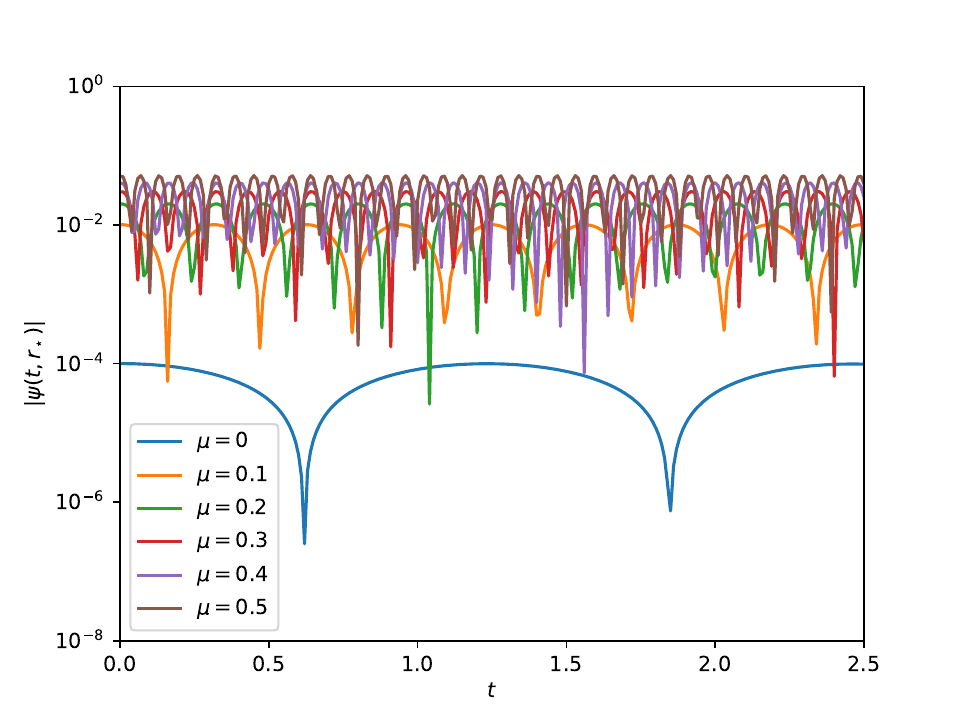}
    \caption{Scalar perturbation as a function of $t$ for $\lambda = 10^{5}$, $\mathcal{Q}=0.1$, $\Lambda = 10^{-5}$, $\alpha = 1$, $\mathcal{M}_{\alpha}=1$, $\ell = 1$ and $q = 0.1$.}
    \label{figscpertmu}
\end{figure}

\begin{figure}
    \centering
    \includegraphics[scale=0.45]{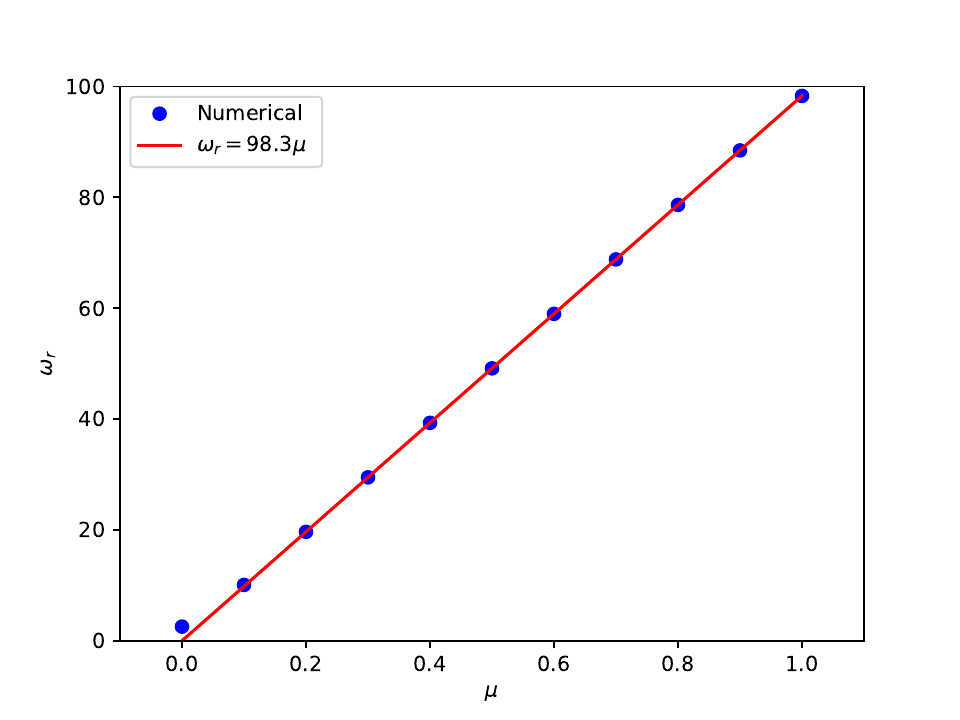}
    \caption{Real part of the frequencies shown in Fig. \ref{figscpertmu} }
    \label{figfreqrmu}
\end{figure}

\begin{figure}
    \centering
    \includegraphics[scale=0.45]{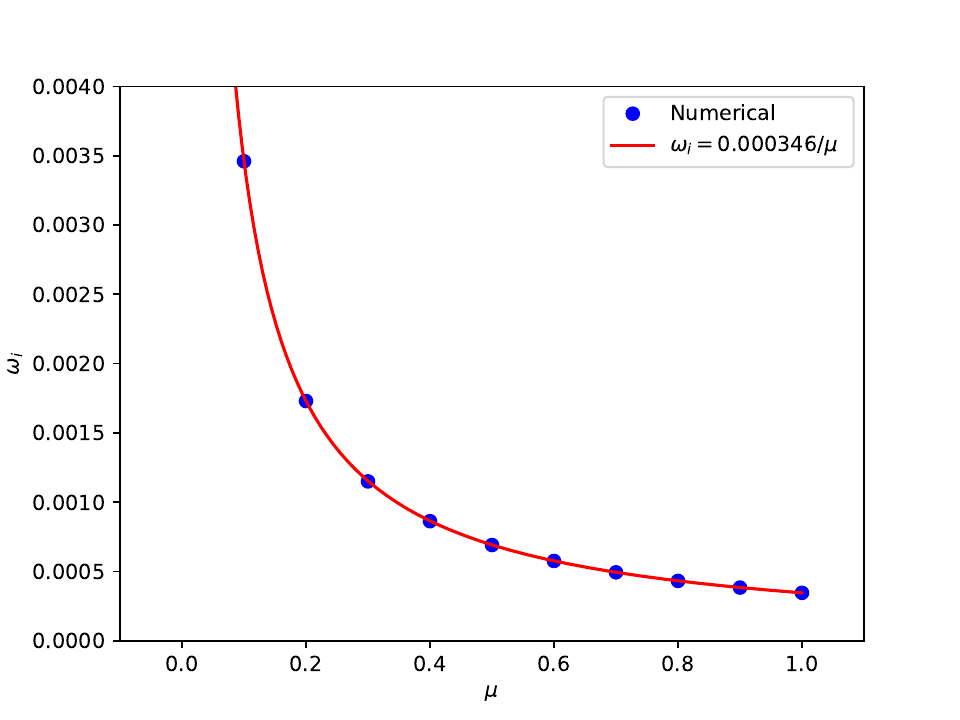}
    \caption{Imaginary part of the frequencies shown in Fig. \ref{figscpertmu} }
    \label{figfreqimu}
\end{figure}

Choosing $\ell = 1$ and different values of $\mu$ we get the behavior shown in Fig. \ref{figscpertmu}. Here both real and imaginary part of the frequencies can be calculated and can be approximated by $\omega = A \mu + i \frac{B}{\mu}$ where
$A = 98.3$ and $B = 0.000346$. These approximations are shown in Figs \ref{figfreqrmu} and \ref{figfreqimu}.

So far we did not find unstable modes and the real frequencies we calculated do not belong to the range given by Eq.(\ref{src}), thus, the modes shown along this section are not superradiant.


\subsection{Electromagnetic perturbations}

In this section we put forward the analysis of the propagation of a test electromagnetic field in the background of a large black hole. In order to achieve this task, we revisit the wave equations governing a test electromagnetic field as follows,
\begin{equation}
\frac{1}{\sqrt{-g}}\partial_{\nu}\left[ \sqrt{-g} g^{\alpha \mu}g^{\sigma \nu} \left(A_{\sigma,  \alpha} -A_{\alpha,  \sigma}\right) \right]=0,  
\end{equation}
where the four-potential, labeled as $A_{\mu}$, can be expressed through an expansion in 4-dimensional vector spherical harmonics in the following fashion
\cite{Toshmatov:2017bpx},  
\begin{small}
\begin{align}\notag
& A_{\mu }\left( t,  r,  \theta, \phi \right)  \nonumber \\
&=\sum_{l, m} 
\begin{bmatrix} 
f(t,  r)Y_{l m}\left( \theta, \phi \right) \\
h(t,  r)Y_{l m}\left( \theta, \phi \right) \\
\frac{a(t,  r)}{\sin \left( \theta \right) }\partial _{\phi }Y_{l
m}\left( \theta, \phi \right) + k(t,  r)\partial _{\theta }Y_{l m}\left( \theta, \phi \right)\\
-a\left( t,  r\right) \sin \left( \theta \right) \partial _{\theta }Y_{lm}\left( \theta, \phi \right)+ k(t,  r)\partial _{\varphi }Y_{l m}\left( \theta, \phi \right)
\end{bmatrix},%
\end{align}%
\end{small}
here $Y_{l m}(\theta,  \phi)$ represents the spherical harmonics in this expansion. The first term on the right-hand side exhibits a parity  $\left(-1\right)^{l +1}$ and is commonly known as the axial sector, meanwhile the second term carries a parity  $\left(-1\right)^l$ referred as the polar sector. Upon directly substituting this expansion into the Maxwell equations, a second-order differential equation for the radial component emerges (detailed in \cite{Toshmatov:2017bpx}),
\begin{equation}
\frac{\mathrm{d}^{2}\Psi \left( r_{\ast }\right) }{\mathrm{d}r_{\ast }^{2}}+\left[ \omega
^{2}-V_{E}\left( r_{\ast }\right) \right] \Psi \left( r_{\ast }\right) =0.
\end{equation}%

In both axial and polar sectors we derive a second-order differential equation governing the radial component by means of the relation $r_{\ast} = \int f^{-1}(r)\mathrm{d}r$ which denotes the tortoise coordinate. The mode $\Psi(r_{\ast})$ is a composite function involving $a(t,  r)$, $f(t,  r)$, $h(t,  r)$, and $k(t,  r)$. However, the precise functional relations differ based on the parity. In the axial sector the mode is articulated as, 
\begin{equation}
a(t,  r)=\Psi \left( r_{\ast }\right),  
\end{equation}
and for the polar sector we have,
\begin{equation}
\Psi \left( r_{\ast }\right) =\frac{r^{2}}{%
l (l +1)}\left[ \partial _{t}h(t,  r)-\partial _{r}f(t,  r)\right].
\end{equation}

In our specific scenario the corresponding effective potential turns to be,
\begin{eqnarray}
    V_{eff}(r)=f(r) \left(\frac{l(l+1)}{r^2}\right),  
\end{eqnarray}
with $f(r)$ given in Eq.(\ref{metric16}).

The behavior of the vector perturbations is displayed in Tables \ref{TabIV} and \ref{TabV}, respectively. With the increase of charge the values of the real part of the quasinormal frequencies increase.

\begin{table}[!h]
\begin{center}
\caption{\label{TabV} By employing the sixth-order WKB approximation, we calculate the quasinormal frequencies for vectorial perturbations related to various values of $\mathcal{Q}$, when $\mathcal{M}_{\alpha} =1$,  $\lambda=10^{5}$, $\Lambda=10^{-5}$, $\alpha =1$, and multipole number  $l=1$.}
\begin{tabular}{c| c | c | c} 
 \hline\hline\hline 
  $\mathcal{Q}$    & $\omega_{0}$ & $\omega_{1}$ & $\omega_{2}$  \\ [0.2ex] 
 \hline 
 0.00  & 0.24818 - 0.09263$i$  & 0.21428 - 0.29410$i$  &  0.17398 - 0.53003$i$  \\
 
 0.10  & 0.24746 - 0.09236$i$ &  0.21366 - 0.29325$i$ &  0.17348 - 0.52849$i$  \\ 
 
 0.20  & 0.25005 - 0.09216$i$ & 0.21723 - 0.29216$i$ &  0.17834 - 0.52521$i$ \\
 
 0.30  & 0.25657 - 0.09186$i$ & 0.22582 - 0.29026$i$ &  0.18961 - 0.51899$i$ \\
 
 0.40 & 0.25635 - 0.09042$i$ & 0.22702 - 0.28521$i$ & 0.19257 - 0.50865$i$ \\
 
 0.50 & 0.25440 - 0.08843$i$ & 0.22657 - 0.27850$i$ & 0.19393 - 0.49547$i$  \\
   [0.2ex] 
 \hline \hline \hline 
\end{tabular}
\end{center}
\end{table}

\begin{table}[!h]
\begin{center}
\caption{\label{TabIV} By using the sixth--order WKB approximation, we investigate the quasinormal frequencies for vectorial perturbations linked to various values of $\mathcal{Q}$, when $\mathcal{M}_{\alpha} =1$,  $\lambda=10^{5}$, $\Lambda=10^{-5}$, $\alpha =1$, and multipole number $l=2$.}
\begin{tabular}{c| c | c | c} 
 \hline\hline\hline 
  $\mathcal{Q}$    & $\omega_{0}$ & $\omega_{1}$ & $\omega_{2}$  \\ [0.2ex] 
 \hline 
 0.00  & 0.45757 - 0.09500$i$ &  0.43651 - 0.29071$i$   &  0.40089 - 0.50170$i$  \\ 
 
 0.10  & 0.46584 - 0.09555$i$ & 0.44522 - 0.29222$i$  & 0.41034 - 0.50380$i$ \\
 
 0.20  & 0.46885 - 0.09491$i$ & 0.44893 - 0.29011$i$  &  0.41526 - 0.49965$i$  \\
 
 0.30  &  0.47055 - 0.09394$i$ & 0.45141 - 0.28697$i$ &  0.41907 - 0.49373$i$ \\
 
 0.40 & 0.46929 - 0.09237$i$ & 0.45103 - 0.28203$i$  & 0.42016 - 0.48471$i$ \\
 
 0.50 & 0.46494 - 0.09027$i$ & 0.44760 - 0.27546$i$  & 0.41828 - 0.47295$i$ \\
   [0.2ex] 
 \hline \hline \hline 
\end{tabular}
\end{center}
\end{table}

So far, we can see from our results that our black hole solution is stable under the perturbations considered here because in the decomposition $\omega=\omega_{\Re}-i \omega_{\Im}$, $\omega_{\Im}>0$. In the next section we will analyze the geodesic motion of massless and massive particles in the background BH geometry.
\section{Geodesics}
\label{sectVI}

The exploration of particle motion in the dark matter context has attracted considerable interest given its profound theoretical implications. A focal point of this inquiry involves understanding the geodesic characteristics of a charged BH with a Yukawa potential. This investigation is crucial for unravelling various astrophysical phenomena associated with these celestial entities, such as the nature of accretion disks and shadows. Essentially, it provides a unique vantage to gain insights into the role of dark matter in this specific context. Our focus is precisely a comprehensive examination of the behavior dictated by the geodesic equation. To achieve this ambitious goal, we write the geodesic equations as,
\begin{equation}
\frac{\mathrm{d}^{2}x^{\mu}}{\mathrm{d}s^{2}} + \Gamma\indices{^\mu_\alpha_\beta}\frac{\mathrm{d}x^{\alpha}}{\mathrm{d}s}\frac{\mathrm{d}x^{\beta}}{\mathrm{d}s} = 0. \label{geodesicscompleteq}
\end{equation}
Here, denoting an arbitrary affine parameter as $s$, our exploration gives rise to four coupled partial differential equations. These equations can be expressed as follows,
\begin{equation}
\begin{split}
&\frac{\mathrm{d}}{\mathrm{d}s}t' = \frac{r' t' \left(-\frac{\alpha  \left(\mathcal{M}-\frac{3 \pi  \mathcal{Q}^2}{32 \ell_0}\right)}{(\alpha +1) \lambda ^2}+\frac{2 \mathcal{Q}}{r^3}-\frac{2 \mathcal{M}}{r^2}+\frac{2 \Lambda  r}{3}\right)}{\frac{\alpha  r \left(\mathcal{M}-\frac{3 \pi  \mathcal{Q}^2}{32 \ell_0}\right)}{(\alpha +1) \lambda ^2}-\frac{\Lambda  r^2}{3}+\frac{\mathcal{Q}}{r^2}-\frac{2 \mathcal{M}}{r}+1}, 
\end{split}
\end{equation}

\begin{equation}
\begin{split}
\frac{\mathrm{d}}{\mathrm{d}s}r' = & \frac{96 \left(r'\right)^2 \left(32 \Xi+9 \pi  \alpha  \mathcal{Q}^2 r^3\right)}{192\Tilde{\Xi}} \\
& + \frac{\Tilde{\Xi}\left(\frac{\alpha  \left(\mathcal{M}-\frac{3 \pi  \mathcal{Q}^2}{32 \ell_0}\right)}{(\alpha +1) \lambda ^2}-\frac{2 \mathcal{Q}}{r^3}+\frac{2 \mathcal{M}}{r^2}-\frac{2 \Lambda  r}{3}\right)\left(t'\right)^2}{192(\alpha +1) \lambda ^2 \ell_0 r^2} \\
& - \frac{2 \left(\theta '\right)^2 \Tilde{\Xi}}{192(\alpha +1) \lambda ^2 \ell_0 r}  - \frac{2 \sin ^2(\theta ) \left(\varphi '\right)^2 \Tilde{\Xi}}{192(\alpha +1) \lambda ^2 \ell_0 r},  
\end{split}
\end{equation}

\begin{equation}
\begin{split}
&\frac{\mathrm{d}}{\mathrm{d}s}\theta' = \sin (\theta ) \cos (\theta ) \left(\varphi '\right)^2-\frac{2 \theta ' r'}{r},  
\end{split}
\end{equation}
and
\begin{equation}
\begin{split}
&\frac{\mathrm{d}}{\mathrm{d}s}\varphi' = -\frac{2 \varphi ' \left(r'+r \theta ' \cot (\theta )\right)}{r}\,,
\end{split}
\end{equation}
where we have defined,  
\begin{equation}
\begin{split}
\Xi = & \ell_0 \left(6 (\alpha +1) \lambda ^2 \mathcal{Q}+2 (\alpha +1) \lambda ^2 \Lambda  r^4 \right. \\
& \left. -3 \alpha  r^3 \mathcal{M} -6 (\alpha +1) \lambda ^2 r \mathcal{M}\right), 
\end{split}
\end{equation} 
\ie
\begin{split}
 \Upsilon = & \left(-3 (\alpha +1) \lambda ^2 \mathcal{Q}+(\alpha +1) \lambda ^2 \Lambda  r^4-3 \alpha  r^3 \mathcal{M} \right. \\
 & \ell_0 \left.- 3 (\alpha +1) \lambda ^2 r^2 + 6 (\alpha +1) \lambda ^2 r \mathcal{M}\right),  
\end{split}
\fe
and
\ie
\Tilde{\Xi} = 32  r \Upsilon + 9 \pi  \alpha  \mathcal{Q}^2 r^4.
\fe

\begin{figure}
    \centering
    \includegraphics[scale=0.45]{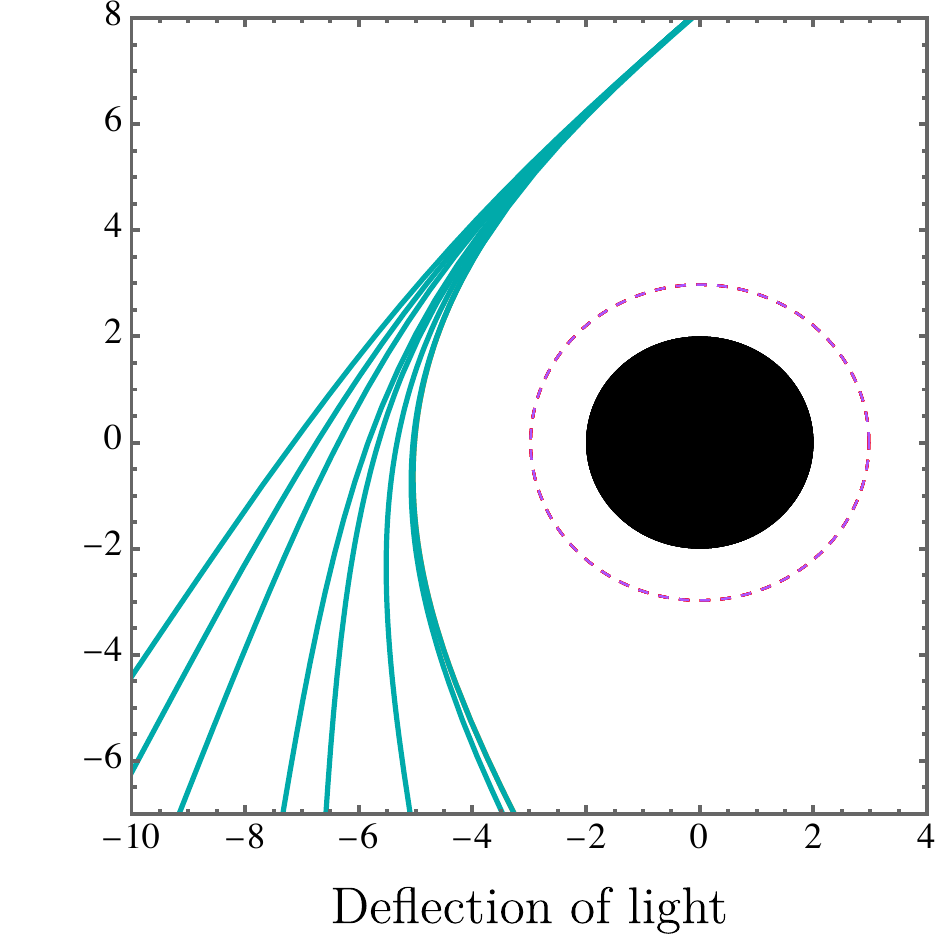}
    \caption{Different light paths for diverse values of $\mathcal{Q}$ when $\ell_0=1$, $\mathcal{M}=1$, $\alpha =1$, $\lambda =10^5$, $\Lambda = 10^{-5}$}
    \label{geodesics}
\end{figure}

\begin{figure}
    \centering
    \includegraphics[scale=0.45]{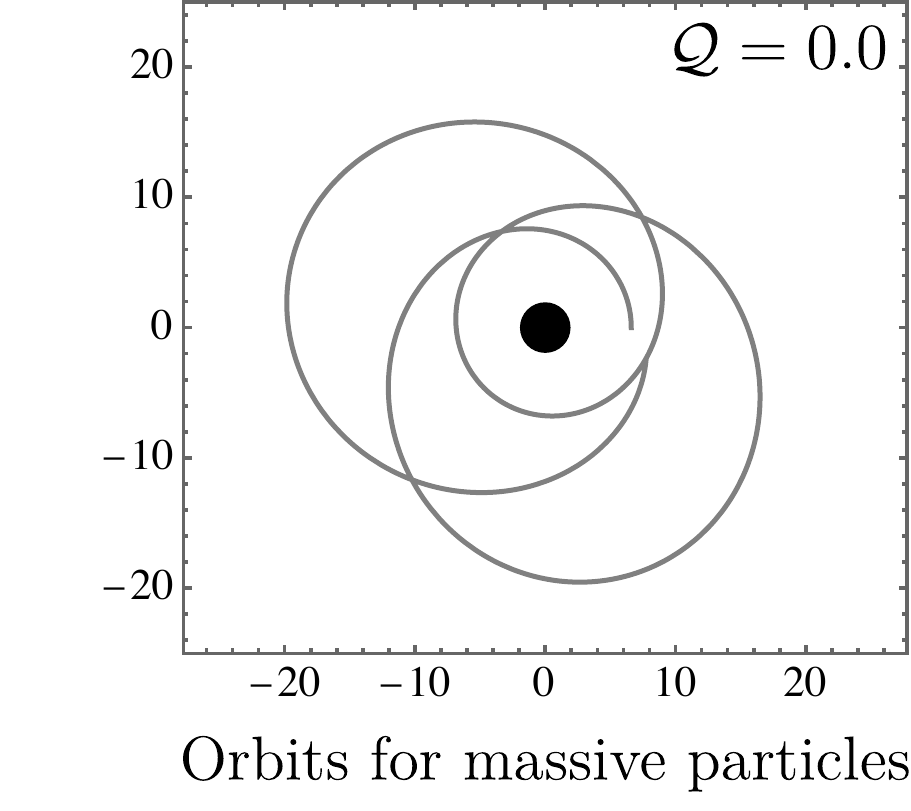}
    \caption{Trajectories of massive particles for diverse values of $\mathcal{Q}$ when $\ell_0=1$, $\mathcal{M}=1$, $\alpha =1$, $\lambda =10^5$, $\Lambda = 10^{-5}$. The black dot represents the event horizon.}
    \label{geodesicsmassive00}
\end{figure}

\begin{figure}
    \centering
    \includegraphics[scale=0.45]{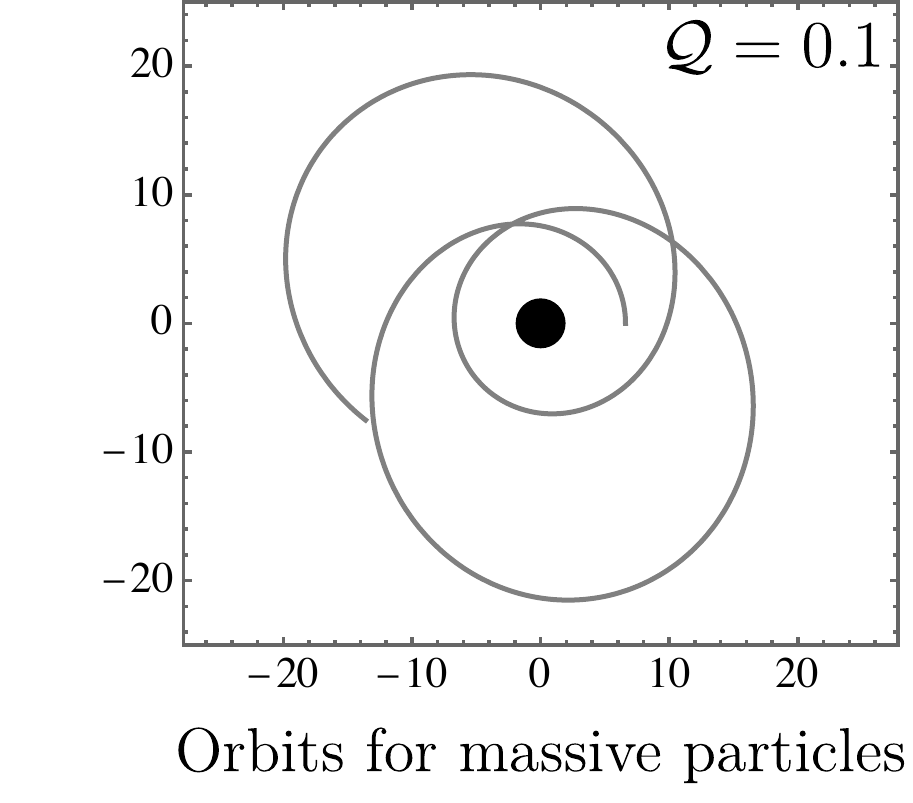}
    \caption{Trajectories of massive particles for diverse values of $\mathcal{Q}$ when $\ell_0=1$, $\mathcal{M}=1$, $\alpha =1$, $\lambda =10^5$, $\Lambda = 10^{-5}$. The black dot represents the event horizon.}
    \label{geodesicsmassive01}
\end{figure}

\begin{figure}
    \centering
    \includegraphics[scale=0.45]{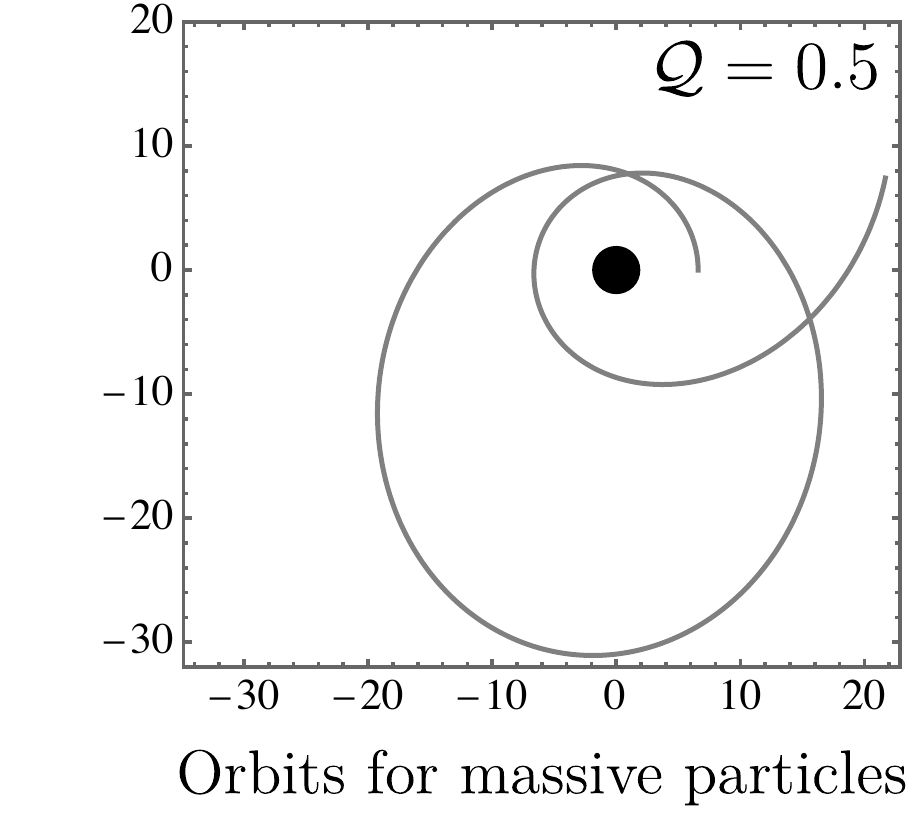}
    \caption{Trajectories of massive particles for diverse values of $\mathcal{Q}$ when $\ell_0=1$, $\mathcal{M}=1$, $\alpha =1$, $\lambda =10^5$, $\Lambda = 10^{-5}$. The black dot again represents the event horizon.}
    \label{geodesicsmassive05}
\end{figure}

In Fig. \ref{geodesics} we display the configuration of the light-like geodesics. Particularly, it shows different light paths for diverse values of $\mathcal{Q}$ in the following order: from the line that is most distant from the BH until the penultimate closest line to it we vary the charge in integer values from $\mathcal{Q}=7$ to $\mathcal{Q}=1$; the closest line to the BH represents $\mathcal{Q}=0.1$. Moreover, these configurations are computed when we keep the other parameters fixed, i.e., $\ell_0=1$, $\mathcal{M}=1$, $\alpha =1$, $\lambda =10^5$, $\Lambda = 10^{-5}$. In addition, we show the orbits of massive particles in Figs. \ref{geodesicsmassive00}, \ref{geodesicsmassive01}, \ref{geodesicsmassive05}. Here, we also consider the same values of $\mathcal{M}, \mathcal{M}_{\alpha}, \lambda, \Lambda$, and $\alpha$ used for the deflection of light. Notice that when $\mathcal{Q}$ increases, a substantial enlargement of trajectories for the massive particles exists.

In the next section we will focus on the lightlike trajectories that give rise to the shadow of a black hole. 

\section{Black hole shadow }
\label{sectVII}

A thorough grasp of critical orbits is essential to understand the complex dynamics of particles and the behavior of light rays near BH structures. In our study these orbits play a pivotal role in unravelling the intricacies of spacetime influenced by the effects of dark matter.

\begin{figure*}
\centering
\includegraphics[width=8.3cm]{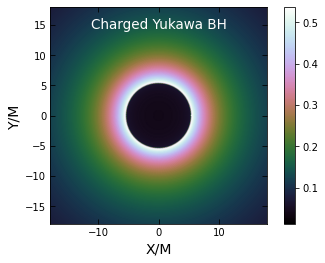}
\includegraphics[width=8.3cm]{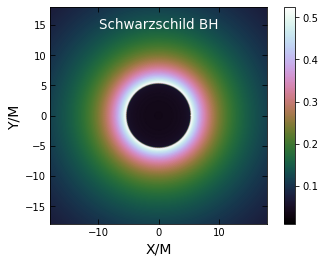}
\caption{The plot for the shadow image of Yukawa charged BH and Schwarzschild BH. For the Yukawa charged BH we have used $\alpha =1, \mathcal{Q}=0.5, \lambda =
10^{10}, \Lambda = 10^{-52}$, and $\mathcal{Q}=0$ for the Schwarzschild BH, respectively. }\label{shadow}
\end{figure*}

To gain a deeper understanding of the impact on the photon sphere, also known as the critical orbit around the BH, we will employ the Lagrangian method to calculate null geodesics. This method offers a more accessible approach for readers to comprehend the calculations compared to those using the geodesic equation, as presented previously. By analyzing the influence of the BH mass on the photon sphere, we aim to extract additional information regarding the gravitational effects of a charged BH with a Yukawa potential. These findings may have significant implications for future observational astronomy. Thus, we begin by considering the following Lagrangian,
\begin{equation}
\mathcal{L} = \frac{1}{2} g_{\mu\nu}\Dot{x}^{\mu}\Dot{x}^{\nu},  
\end{equation}
when contemplating a constant angle of \(\theta = \pi/2\), the previously mentioned expression simplifies to,
\ie
g_{00}^{-1} E^{2} + g_{11}^{-1} \Dot{r}^{2} + g_{33}L^{2} = 0,  
\fe
with \(E\) representing the energy and \(L\) denoting the angular momentum of the test particle. Rearranging the previous expression we can arrive at the radial equation,
\ie
\Dot{r}^{2} = E^{2} - \mathcal{V},  
\label{motion}
\fe
where $\mathcal{V} \equiv \left(  -\frac{\Lambda  r^2}{3}+\frac{\mathcal{Q}}{r^2}-\frac{2 \mathcal{M}}{r}+\frac{r \mathcal{M}_{\alpha}}{(\alpha +1) \lambda ^2}+1 \right)\left(  \frac{L^{2}}{r^{2}} \right)$ is the effective potential. In the quest to determine the critical radius, we should solve the equation $\frac{\partial \mathcal{V}}{\partial r} = 0$. 
Only one of the three roots that this equation possesses emerges as physically meaningful since it is real and positive and it is known as the photon sphere radius,
\ie
\begin{split}
r_{c} =& -\frac{2 \left(\alpha  \lambda ^2+\lambda ^2\right)}{3 \mathcal{M}_{\alpha }} + \frac{\Omega}{3 \mathcal{M}_{\alpha} \Tilde{\rho}}  -\frac{\Tilde{\mathcal{\rho}}}{3 \sqrt[3]{2} \mathcal{M}_{\alpha }},  
\end{split}
\fe
where
\ie
\Omega =\sqrt[3]{2} \left(-4 \left(\alpha  \lambda ^2+\lambda ^2\right)^2-18 \mathcal{M}_{\alpha } \left(\alpha  \lambda ^2 \mathcal{M}+\lambda ^2 \mathcal{M}\right)\right),  
\fe
and
\ie
\begin{split}
& \Tilde{\rho} = \,  \left(48 \alpha  \lambda ^6+16 \lambda ^6 + 16 \alpha ^3 \lambda ^6 + 48 \alpha ^2 \lambda ^6 \right. \\
& \left. + 108 \lambda ^4 \mathcal{M} \mathcal{M}_{\alpha }   108 \alpha ^2 \lambda ^4 \mathcal{M} \mathcal{M}_{\alpha }+216 \alpha  \lambda ^4 \mathcal{M} \mathcal{M}_{\alpha } \right. \\
&\left. +108 \lambda ^2 \mathcal{Q} \mathcal{M}_{\alpha }^2 +108 \alpha  \lambda ^2 \mathcal{Q} \mathcal{M}_{\alpha }^2  \right. \\
& \left. + \left( 4 \left(-4 \left(\alpha  \lambda ^2+\lambda ^2\right)^2-18 \mathcal{M}_{\alpha } \left(\alpha  \lambda ^2 \mathcal{M}+\lambda ^2 \mathcal{M}\right)\right){}^3 \right.\right. \\
& \left.\left. \left( 16 \alpha ^3 \lambda ^6+48 \alpha ^2 \lambda ^6+48 \alpha  \lambda ^6+16 \lambda ^6  \right.\right.\right. \\
& \left.\left.\left. +  108 \alpha ^2 \lambda ^4 \mathcal{M} \mathcal{M}_{\alpha }+108 \lambda ^4 \mathcal{M} \mathcal{M}_{\alpha }+216 \alpha  \lambda ^4 \mathcal{M} \mathcal{M}_{\alpha }   \right.\right.\right. \\
& \left.\left.\left. +108 \lambda ^2 \mathcal{Q} \mathcal{M}_{\alpha }^2+108 \alpha  \lambda ^2 \mathcal{Q} \mathcal{M}_{\alpha }^2        \right)^{2}                     \right)^{1/2}              \right)^{1/3}.
\end{split}
\fe

Next, in terms of the photon sphere radius we can express the shadow radius of the BH as,
\ie
\begin{split}
& \tilde{\mathcal{R}}  = \frac{r_{c}}{{\sqrt{f(r_{c})}}} \\
& = \frac{4 (\alpha +1) \lambda ^2 \Tilde{\rho} - \Tilde{\rho}^2 2^{2/3} +2 \Omega }{2 \Tilde{\rho}  \mathcal{M}_{\alpha } \sqrt{-\frac{3 \left(-3 \mathcal{Q}+\Lambda  r^4+6 r \mathcal{M}\right)}{r^2}+\frac{9 r \mathcal{M}_{\alpha }}{(\alpha +1) \lambda ^2}+9}}.
\end{split}
\fe
In the general case, as was shown in \cite{Filho:2023abd}, the location of the observer should be taken into account, namely,  $\tilde{\mathcal{R}}= r_c\sqrt{f(r_O)/f(r_{c})})$. However, we can neglect the effect of $r_O$ for astrophysical values of parameters since $f(r_O)\to 1$. In Fig. \ref{shadow} we have plotted the shadow images of the Yukawa BH and the Schwarzschild BH using the infalling gas model described in what follows. 

We considered a $4$-velocity given by,
\begin{eqnarray}
u^{\mu}=\left(\frac{1}{f(r)},  -\sqrt{1-f(r)},  0, 0\right),  
\end{eqnarray}
with an observed specific intensity $I_{\nu 0}$ at the observed photon frequency $\nu_\text{obs}$ and at the point $(X,  Y)$ of the observer's image  determined by \cite{Bambi:2013nla,Nampalliwar:2021tyz,Jusufi:2021lei,Saurabh:2020zqg},
\begin{eqnarray}
    I_{obs}(\nu_{obs},  X,  Y) = \int_{\gamma}\mathrm{g}^3 j(\nu_{e})dl_\text{prop}, 
\end{eqnarray}
where the redshift function $\mathrm{g}$ is specified by, 
\begin{eqnarray}
  \mathrm{g} =\frac{p_{\mu}u_{obs}^{\mu}}{p_{\nu}u_e^{\nu}},
\end{eqnarray}
here $u_{obs}^{\mu}$ gives the $4$-velocity of the 
observer, i.e.,  $u_{obs}^{\mu}=(1,0,0,0)$ and $p^{\mu}$ is the photon $4$-momentum. Note also that in the above equations we have the proper length $dl_\text{prop}$ and an inverse square law, i.e., $1/r^2$ for the emission $j(\nu_{e}) \propto 1/r^2$ (for more details about the infalling gas see \cite{Filho:2023abd}). In our plots we used values for $\lambda$ and $\Lambda$ based on astrophysical data. In particular, we take $\lambda \sim 10^{10}$ in units of BH mass and $\Lambda \sim 10^{-52}$ m$^{-2}$. In principle, we can distinguish the charged Yukawa BH only from the Schwarzschild BH, but it is tough to distinguish it from the Reissner-Nordstrom BH. For a specific value $\mathcal{Q}=0.5$ the shadow radius of the charged Yukawa BH is $\tilde{\mathcal{R}} \simeq 4.968$ [in units of BH mass] which is smaller compared to the Schwarzschild BH which is $\tilde{\mathcal{R}} \simeq 5.196$ [in units of BH mass]. 


In the next section we will make an exhaustive study of the process of accretion on the charged Yukawa BH.

\section{Accretion of matter on BH}
\label{sectVIII}

Babichev et al. \cite{Babichev:2013vji} studied the basic dynamical equations of
dark energy interacting with BHs, which cover a wide range
of field theory models, as well as perfect fluids with various
equations of state, including cosmological dark energy. For astrophysical BHs, as we pointed out, we can neglect the effect of $\ell_0$. Hence, we shall explore the implications of the metric \eqref{metric16}. For our
current work we take a perfect fluid whose stress-energy tensor
is as follows \cite{Jawad:2021hay, Jawad:2016fop, Jawad:2017zkt,  Salahshoor:2018plr}
\begin{equation}\label{55}
T^{\mu v}=(p+\rho)u^{\mu}u^{v}+pg^{\mu v},  
\end{equation}
with
\begin{equation}\label{56}
u^{\mu}=\frac{\mathrm{d}x^{\mu}}{\mathrm{d}\tau}=(u^{t},  u^{r},  0,  0),  
\end{equation}
where $\tau$ shows the proper time along the geodesic motion. Since
our fluid is in steady state, spherically symmetric, and fulfils
normalization condition $u^{\mu}u_{\mu}=-1$, one can obtain $u^{t}=1/f(r)$ and $u^r=\pm \sqrt{1-f(r)}$, where in $u^r$ the negative or positive sign describes the ingoing and outgoing particles. In general, we can relate to them
\begin{eqnarray}
    u^t=\sqrt{\frac{f(r)+(u^r)^2}{f(r)}}.
\end{eqnarray}
The conservation of stress-energy tensor is as follows
\begin{equation}\label{59}
T^{\mu v}_{;\mu}=0\Rightarrow T^{\mu
v}_{;\mu}=\frac{1}{\sqrt{-g}}(\sqrt{-g}T^{\mu
v})_{,  \mu}+\Gamma^{v}_{a\mu}T^{a \mu}=0,  
\end{equation}
and after a few manipulations, we get
\begin{equation}\label{62}
(p+\rho)u^{r}r^{2}\sqrt{(u^{r})^{2}+f(r)}=B_{0},  
\end{equation}
where $B_{0}$ is a constant of integration. Applying stress-energy
conservation law on four velocities, after some simplifications, we get
\begin{equation}\label{66}
\frac{\rho'}{(p+\rho)}+\frac{u'}{u}+\frac{2}{r}=0,  
\end{equation}
and after integration, it will be
\begin{equation}\label{67}
r^{2}u^{r}\exp\left(\int\frac{\mathrm{d}\rho}{p+\rho}\right)=-B_{1},  
\end{equation}
where $B_{1}$ is an integration constant. Since $u^{r}<0$, the right-hand side also
takes a minus sign. So we have
\begin{equation}\label{68}
(p+\rho)\sqrt{f(r)\left[\frac{(u^{r})^{2}}{f(r)}+1\right]}\exp\left(-\int\frac{\mathrm{d}\rho}{p+\rho}\right)=B_{2},  
\end{equation}
being $B_{2}$ an integration constant. The mass flux equation from the above setup is given by,
\begin{equation}\label{69}
(\rho u^{\mu})_{;}\equiv\frac{1}{\sqrt{-g}}(\sqrt{-g}\rho
u^{\mu})_{,  \mu}=0.
\end{equation}
We can also rewrite it as,
\begin{equation}\label{70}
\frac{1}{\sqrt{-g}}(\sqrt{-g}\rho u^{\mu})_{,  r}+
\frac{1}{\sqrt{-g}}(\sqrt{-g}\rho u^{\theta})_{,  \theta}=0.
\end{equation}
As we are restricting our study only to the equatorial plane, then
the second term of Eq. \eqref{70} would vanish, and we obtain the
mass conservation energy equation as
\begin{equation}\label{71}
r^{2}u^{r}{\rho}=B_{3},  
\end{equation}
where $ B_{3}$ is an integration constant.  Then, Eqs. \eqref{67},  
\eqref{68} and \eqref{71} yield,
\begin{equation}\label{72}
\frac{p+\rho}{\rho}\sqrt{f(r)\left[\frac{(u^{r})^{2}}{f(r)}+1\right]}=B_{4},  
\end{equation}
\begin{figure}
\centering
\includegraphics[width=7.2cm]{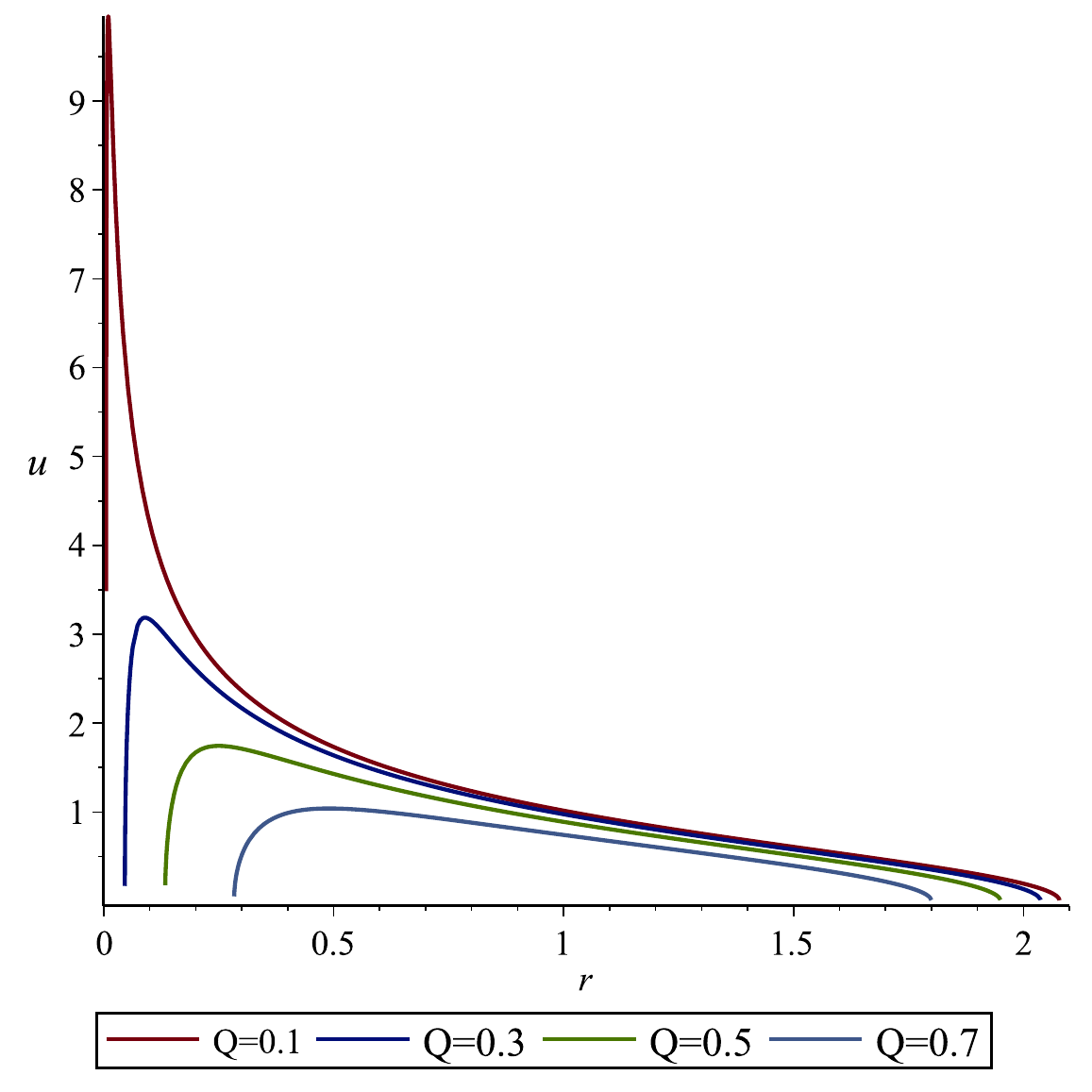}
\includegraphics[width=7.2cm]{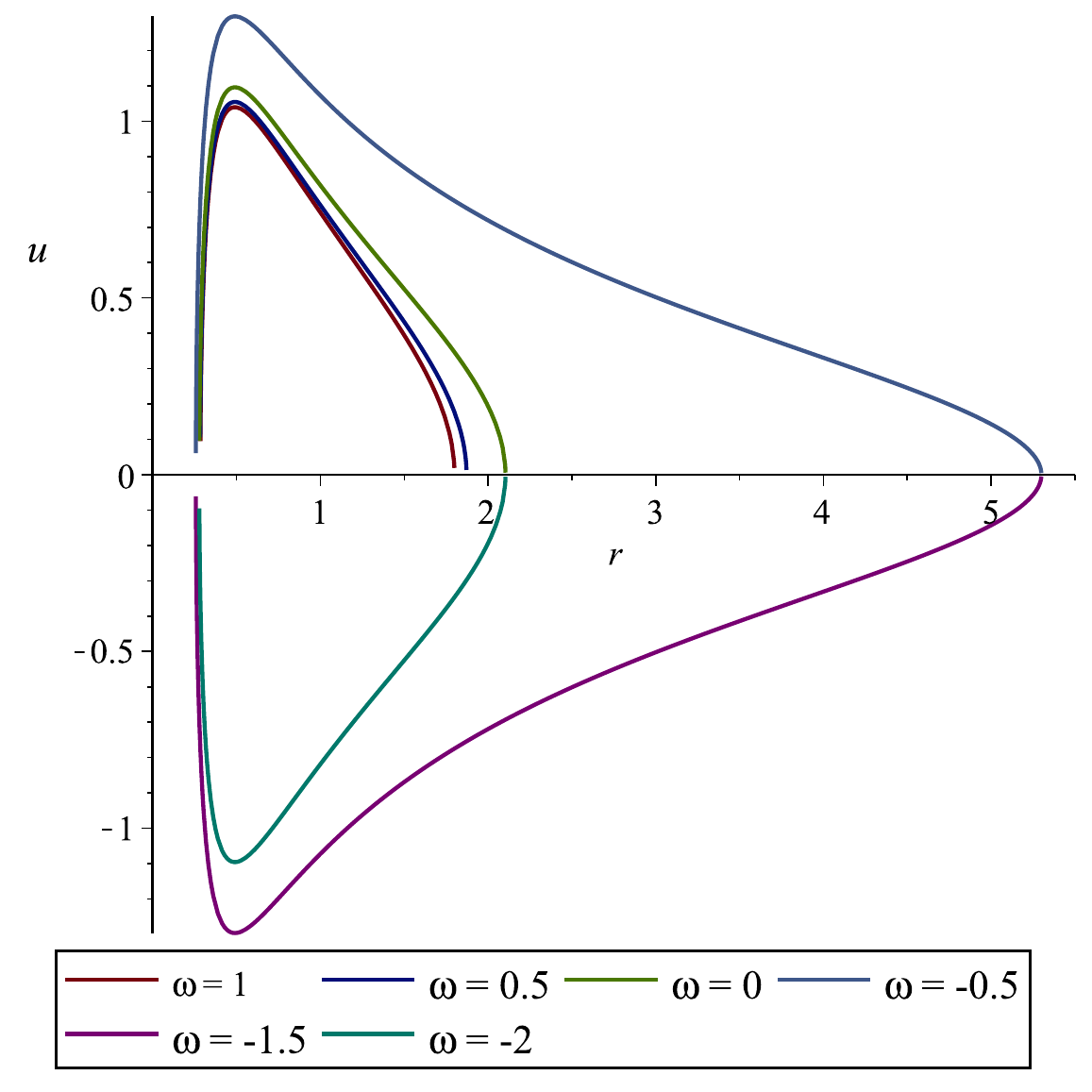}
\caption{Radial velocity for $\omega=1$ (upper panel) and radial
velocity for $\mathcal{Q}=0.7$ (lower panel), where $\alpha =0.5, \lambda =
10^5, \Lambda = 10^{-5}, \ell_0 = 10^{-5}, \omega=1  $.}\label{fig1}
\end{figure}
\begin{figure}
\centering
\includegraphics[width=7.2cm]{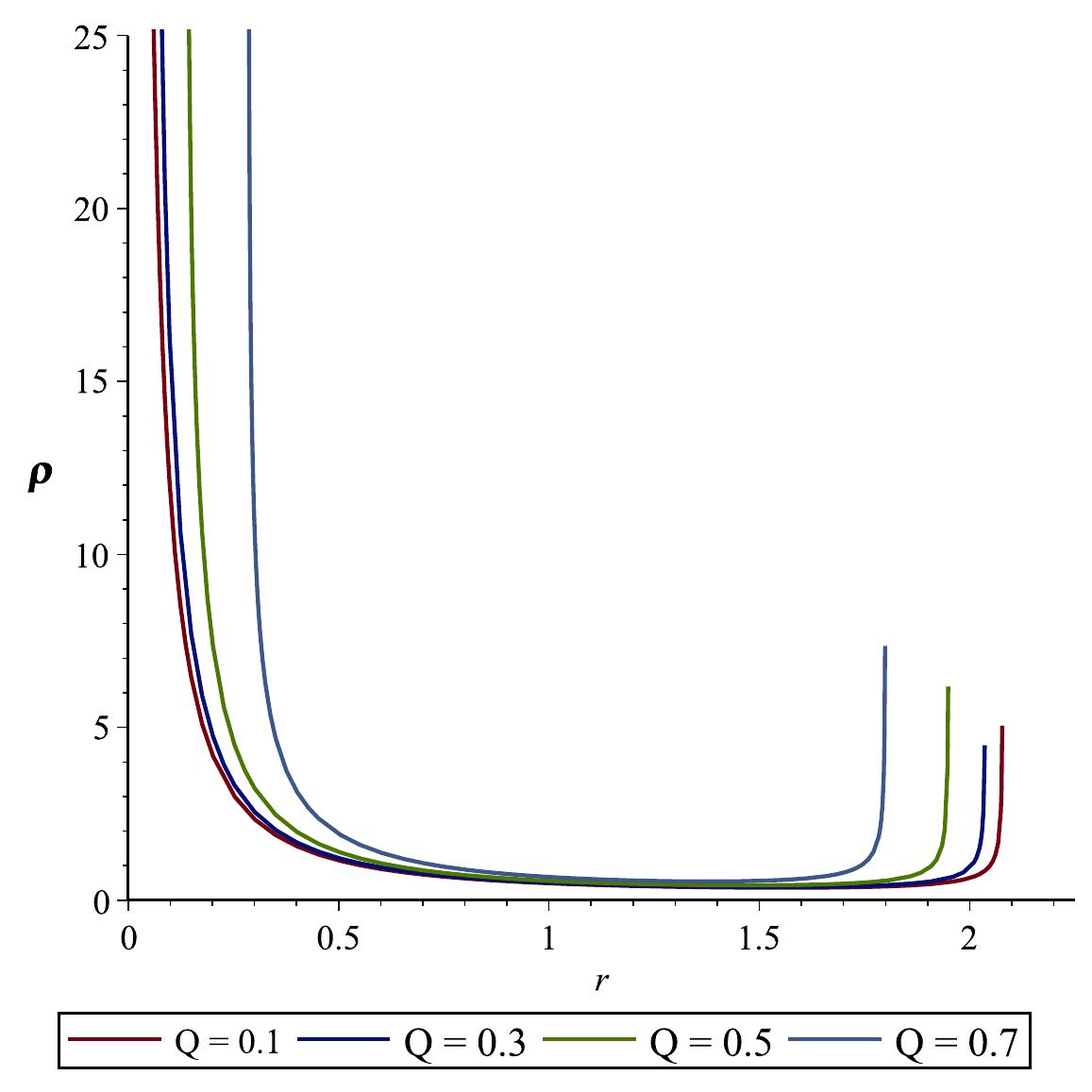}
\includegraphics[width=7.2cm]{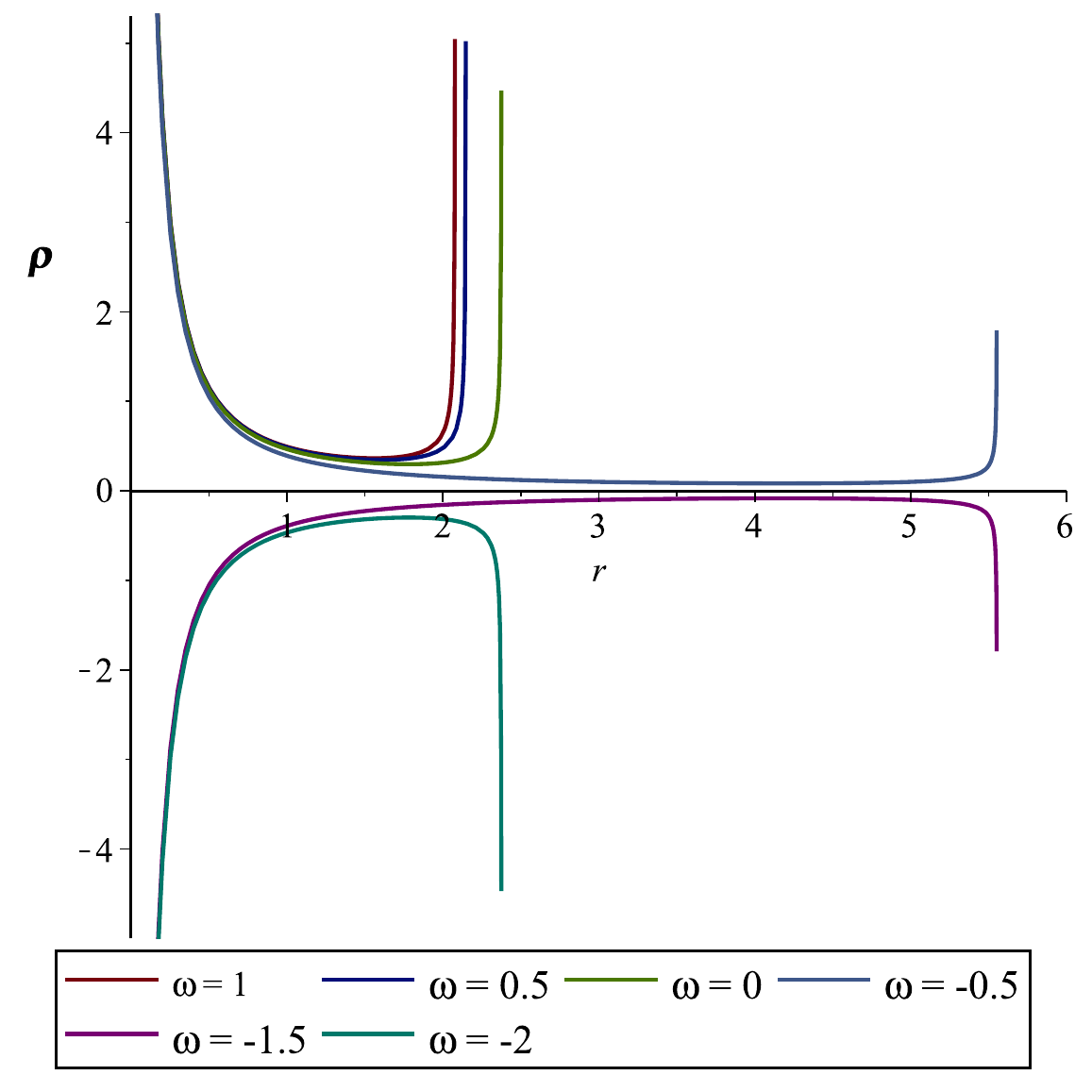}
\caption{Energy density for $\omega=1$ (upper panel) and energy
density for $\mathcal{Q}=0.1$ (lower panel), where $\alpha =0.5, \lambda =
10^5, \Lambda = 10^{-5}, \ell_0 = 10^{-5}, \omega=1$.}\label{fig2}
\end{figure}
being $B_{4}$ another integration constant. Hence, by utilizing the
equation of state $p=\omega\rho$, we get
\begin{equation}\label{73}
u=\left(\frac{1}{\omega+1}\right)\sqrt{f(r)\left[\frac{B^{2}_{4}}{f(r)}-(\omega+1)^{2}\right]}.
\end{equation}
Now, we can obtain the density of the fluid from Eq. \eqref{71} as,
\begin{equation}\label{74}
\rho=\frac{B_{3}}{r^{2}}\frac{(\omega+1)}{\sqrt{B^{2}_{4}-(\omega+1)^{2}f(r)}}.
\end{equation}
Pressure can be found using $p=\omega \rho$ and the above equation.
The mass of a BH is not fixed but depends on the surrounding
environment and the type of dark energy that fills the universe. The
rate of change of mass of a BH can be calculated by measuring how
much matter and energy cross its event horizon in both directions.
This rate is usually denoted by $\dot{\mathcal{M}}$ and it can be expressed
by a mathematical formula that involves the flux of the fluid over
the surface of the BH. We can obtain $\dot{\mathcal{M}}$ as follows \cite{Jawad:2021hay},
\begin{equation}\label{82}
\dot{\mathcal{M}}=4\pi B_2 \mathcal{M}^2 (p+\rho).
\end{equation}

\begin{figure*}
\centering
\includegraphics[width=7.2cm]{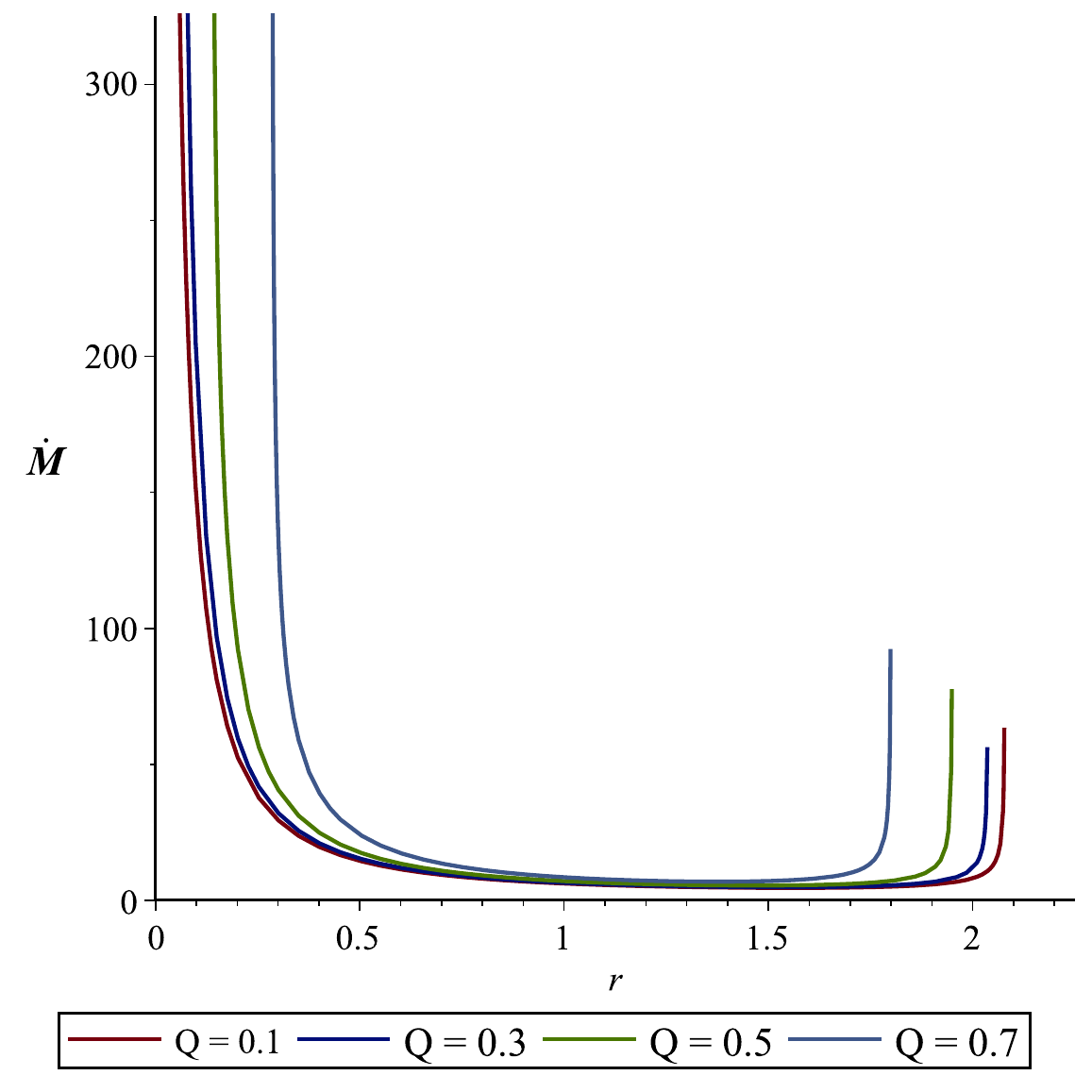}
\includegraphics[width=7.2cm]{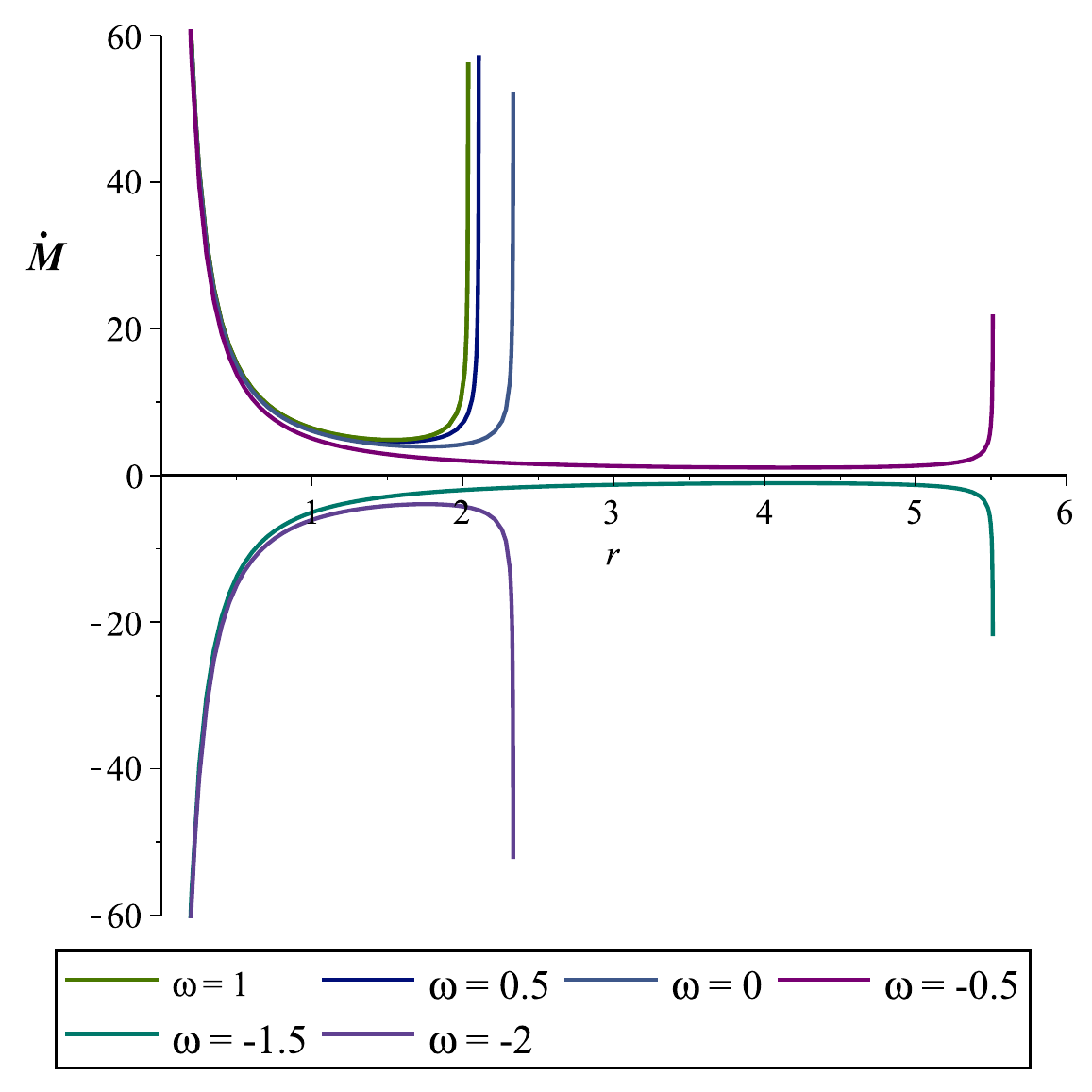}\\
\caption{Rate of change of mass for $\omega=1$ (left panel) and rate
of change of mass for $\mathcal{Q}=0.1$ (right panel), where $\alpha = 0.5,  
\lambda = 10^5, \Lambda = 10^{-5}, \ell_0 = 10^{-5}, \omega=1$.}
\label{fig3}
\end{figure*}

The radial velocity profile is shown in Fig. \ref{fig1} for
different values of $Q$ and $\omega$. The parameter $\omega$ is
related to the type of dark energy. The density profile shows how the fluid's density changes with the
distance from the BH. The density profile is shown in Fig.
\ref{fig2} for different values of $Q$ and $\omega$. The
constants $B_3$ and $B_4$ are arbitrary constants that affect the
shape of the profiles and we have chosen them to be $0.4$ and
$0.5$, respectively. In the upper panel of Fig. \ref{fig1},  
we can see that for a fixed value of $\omega$ increasing $Q$ makes
the radial velocity lower. This means that higher values of $Q$ make
the dark energy stronger, which slows down the fluid’s acceleration.
In the lower panel we can see that for a fixed value of $Q$,  
changing $\omega$ makes the radial velocity positive or negative.
This means that different types of dark energy have different
effects on the fluid’s direction. In the upper panel of Fig.
\ref{fig2} we can see that for a fixed value of $\omega$  
increasing $\mathcal{Q}$ makes the density lower. Higher
values of $\mathcal{Q}$ strengthen the dark energy, reducing the
fluid’s compression. In the lower panel we can see that for a fixed
value of $\mathcal{Q}$ varying $\omega$ makes the density positive or
negative. This means that different types of dark energy have
different effects on the fluid’s mass. We plotted the accretion rate in Fig. \ref{fig3}. In the vicinity of a BH the accretion rate is
getting lower due to a strong effect of gravitation;
as the value of $\mathcal{M}$  decreases, the accretion rate also enhances.

\section{Conclusions}
\label{sectIX}
In the present paper  we obtained a family of charged BH solutions with short and long-range modifications using the Yukawa-like form of the gravitational potential. We can generally include the deformed quantum parameter $\ell_0$ to obtain nonsingular gravitational/electric potential. We find that the mass of the BH gets corrected due to the apparent dark matter mass and the self-corrected electric charge mass. We obtained two specific solutions: a regular BH solution that describes the geometry of small BHs where the quantum effects are expected to be essential and a BH solution that represents the geometry outside large BHs at some considerable distances where the Yukawa corrections are crucial.\\

For the QNMs we found that electric charge increases the value of the real part of QNMs, while the effect of Yukawa parameters is very small. The main impact from Yukawa parameters is encoded in $\alpha$, which is absorbed in the total mass of the BH. 

Furthermore, the thermodynamics analyses show that the regular BH solution can undergo a phase transition at the maximal temperature. For such a BH the quantum effects encoded in $\ell_0$ are expected to play an essential role in the final state. In particular, from the Hawking temperature plot it can be seen that the Hawking temperature vanishes at some critical $r_+$ that should correspond to the extremal BH. This means we can get the so-called BH remnants. From the point of view of a distant observer, in the case of supermassive BHs the effect of $\ell_0$ can be practically neglected. Hence, one can study the BH thermodynamics by ignoring the effect of $\ell_0$. The impact of $\ell_0$ is thus meaningful only for small BHs, not astrophysical ones. 

From the analyses of the shadow images, we argued that one can distinguish the charged Yukawa BH from the Schwarzschild BH but not from the Reissner–Nordstrom. This has to do with the fact that the Yukawa parameters have a negligible effect on the shadow radius. The only significant impact comes from $\alpha$, which can be absorbed into the redefinition of the BH mass. 
 
The radial velocity and density profiles for various values of $\omega$ and $Q$ are presented to analyze how the fluid's density changes with the distance from the BH for different types of dark energy. Also, the arbitrary constants $B_3$ and $B_4$ that affect the shape of the profiles are chosen to be $0.4$ and $0.5$, respectively. We found that higher values of electric charge make the dark energy stronger, indicating that the fluid's acceleration slows. Also, we discovered that different types of dark energy have various effects on the fluid’s direction as the radial velocity becomes positive or negative depending on changing dark energy types. The compression effect of energy density is reduced for higher values of electric charge, while different types of dark energy have various effects on the fluid’s mass. Also, in the vicinity of a BH the accretion rate gets lower due to the strong effect of gravitation. This rate shows the dependence on the $\mathcal{M}$ values as the accretion rate strengthens for its decreasing values.

\bigskip

\begin{acknowledgements}
    A. A. Araújo Filho is supported by Conselho Nacional de Desenvolvimento Cient\'{\i}fico e Tecnol\'{o}gico (CNPq) and Fundação de Apoio à Pesquisa do Estado da Paraíba (FAPESQ) - [200486/2022-5] and [150891/2023-7]. Genly Leon was funded by Vicerrectoría de Investigación y Desarrollo Tecnológico (VRIDT) at Universidad Católica del Norte through Resolución VRIDT No. 026/2023 and Resolución VRIDT No. 027/2023 and the support of Núcleo de Investigación Geometría Diferencial y Aplicaciones, Resolución Vridt No. 096/2022. He also acknowledges the financial support of Proyecto de Investigación Pro Fondecyt Regular 2023, Resoluci\'{o}n VRIDT No. 076/2023. 

\end{acknowledgements}

\bibliographystyle{apsrev4-1}
\bibliography{main}

\end{document}